\documentclass[entropy,a4paper,article,accept,moreauthors,pdftex]{Definitions/mdpi} 

\firstpage{1} 
\pubvolume{23}
\issuenum{8}
\articlenumber{970}
\pubyear{2021}
\copyrightyear{2021}
\externaleditor{{Academic Editors: Fabio Aiolli and Mirko Polato}} 
\datereceived{30 June 2021} 
\dateaccepted{22 July 2021} 
\datepublished{28 July 2021} 
\hreflink{https://doi.org/ 10.3390/e23080970} 
\usepackage{tabularx}
\usepackage{siunitx} 
\usepackage{tablefootnote}

\usepackage{xspace}

\usepackage[labelformat=simple]{subcaption}															
										
\DeclareCaptionLabelFormat{subcaptionlabel}{\normalfont(\textbf{#2}\normalfont)}				
\newcommand{\irn}{R.N.\xspace}
\newcommand{\imf}{M.F.D.\xspace}
\newcommand{\ipc}{P.C.\xspace}

\newcommand{\idest}{i.e.,\xspace}
\newcommand{\eg}{e.g.,\xspace}

\newcommand{\secref}[1]{Section \ref{#1}}
\newcommand{\tabref}[1]{Table \ref{#1}}
\newcommand{\figref}[1]{Figure \ref{#1}}
\newcommand{\ceqref}[1]{Equation \eqref{#1}}

\newcommand{\sciexp}[2]{\ensuremath{#1 \cdot 10^{#2}}}

\newcommand{\matsize}[1]{$|#1| \times |#1|$\xspace}
\newcommand{\twomatsize}[2]{$|#1| \times |#2|$\xspace}
\newcommand{\fsize}{\matsize{F}}
\newcommand{\isize}{\matsize{I}}

\newcommand{\rptb}{RP$^3\beta$\xspace}

\usetikzlibrary{arrows}

\definecolor{mplblue}{HTML}{1F78B4}
\definecolor{tomato}{HTML}{FF6347}
\definecolor{limegreen}{HTML}{32CD32}

\usepackage[final]{changes}

\Title{Feature Selection for Recommender Systems with \mbox{Quantum Computing}}

\TitleCitation{Feature Selection for Recommender Systems with Quantum Computing}

\Author{Riccardo Nembrini 
 $^{1,2,}$* \orcidA{}, Maurizio Ferrari Dacrema$^{1}$ \orcidB{} 
 and Paolo Cremonesi$^{1}$ \orcidC{}}

\AuthorNames{Riccardo Nembrini, Maurizio Ferrari Dacrema and Paolo Cremonesi}
\AuthorCitation{Nembrini, R.; Ferrari Dacrema, M.; Cremonesi, P.}

\address{$^{1}$ \quad Politecnico di Milano, Via Ponzio 34/5, 20133 Milano, Italy; maurizio.ferrari@polimi.it (M.F.D.);  	paolo.cremonesi@polimi.it (P.C.)\\
$^{2}$ \quad ContentWise, Via Privata Simone Schiaffino, 11, 20158 Milano, Italy}

\corres{Correspondence: riccardo.nembrini@polimi.it}

\abstract{
    The promise of quantum computing to open new unexplored possibilities in several scientific fields has been long discussed, but until recently the lack of a functional quantum computer has confined this discussion mostly to theoretical algorithmic papers. It was only in the last few years that small but functional quantum computers have become available to the broader research community. One paradigm in particular,{\emph{quantum annealing},} 
 {can be used to sample optimal solutions for a number of} NP-hard optimization problems represented with classical operations research tools, providing an easy access to the potential of this emerging technology.
    One of the tasks that most naturally fits in this mathematical formulation is feature selection. In this paper, we investigate how to design a hybrid feature selection algorithm for recommender systems that leverages the domain knowledge and behavior hidden in the user interactions data.
    We represent the feature selection as an optimization problem and solve it on a real quantum computer, provided by D-Wave. The results indicate that the proposed approach is effective in selecting a limited set of important features and that quantum computers are becoming powerful enough to enter the wider realm of applied science.
}

\begin{document}

\section{Introduction}
\label{sec:introduction}
The revolutionary potential of quantum computing has been known at a theoretical level for decades, but the development of a functional quantum computer (QC) proved long elusive due to remarkable engineering challenges. For this reason, they have been long confined to few highly specialized laboratories and very expensive to use. In recent years QCs are starting to become accessible to the wider research communities, some of them being freely available on the cloud, for example the QCs of D-Wave 
(\url{https://www.dwavesys.com/take-leap} (accessed on 27 July 2021)) and IBM (\url{https://quantum-computing.ibm.com/} (accessed on 27 July 2021)).
There are two main paradigms for a QC. The first one is the {\emph{gate model},} which is analogous to classical computers in that an electronic circuit applies a sequence of transformations on a {\emph{state}} that is quantum rather than classical. The second paradigm is {\emph{adiabatic computing},} which instead relies on the natural tendency of a physical system to evolve towards, and remain into, a state of minimal energy. Although the two paradigms are equivalent, available QCs built according to the adiabatic paradigm are currently less powerful than the gate model, \idest they are nonuniversal, and are called \mbox{\emph{quantum annealers} \cite{AdiabaticQuantumComputingandQuantumAnnealing}.} This paradigm has gain{ed} popularity due to its ability to {sample optimal solutions for optimization problems (is suitable to address both NP-hard and NP-complete problems \cite{10.3389/fphy.2014.00005}) } formulated with known operations research tools and is the paradigm we will use.
{Whether quantum annealing is able to solve such NP-hard problems efficiently and \idest lower their asymptotic complexity is still an open question, but it is believed not to be the case in general. Nonetheless, there is experimental evidence for substantial constant speed-up on some specific problems \cite{PhysRevX.6.031015}. Although more limited than the gate model, which is known to reduce the computational complexity of several interesting problems, a constant speedup can have significant practical implications. Consider as an example the impact of GPUs, that in the last decade have allowed extraordinary advancements in deep learning (and other machine learning tasks in general) thanks to the constant speedup they offer for the training of such models over traditional CPUs.} 

Although the capabilities of these QCs are still limited, the technology is evolving at a rapid pace to the point that it is now possible to use them to solve small but realistic problems.
The future of this technology rests on two important factors: first is the engineering of new and ever more powerful QCs, second is the development of new algorithmic solutions to leverage their potential.
{Several formulations of important problems have been proposed for quantum annealers, \eg graph isomorphism \cite{Tamascelli_2014}, graph partitioning \cite{DBLP:journals/corr/Ushijima-Mwesigwa17}, Restricted Boltzmann Machines for deep learning \cite{DBLP:journals/corr/AdachiH15}, Support Vector Machines \cite{DBLP:journals/cphysics/WillschWRM20} and matrix factorization \cite{10.1371/journal.pone.0206653}.}

In this paper, we want to explore how QCs could be applied to the Recommender Systems field. We therefore select a task which fits well in the mathematical formulation and current capabilities of the D-Wave quantum annealer, feature selection. Selecting the appropriate set of features to be used for classification or predictive models is, in general, an NP-hard problem and a widely researched topic in many research fields \cite{DBLP:journals/cee/ChandrashekarS14,10.1007/978-981-15-5616-6_10}.
Understanding which is the most appropriate feature set is usually a time consuming task that requires significant domain knowledge and it is an important issue to address when a recommender system is deployed.
It is known that recommender systems built by using past user interactions (\idest \emph{collaborative} models) often tend to provide higher recommendation quality than models based on item features (\idest \emph{content-based} models) but there are several scenarios where collaborative recommenders cannot be applied.
We will consider in particular the \emph{cold start} scenario, where a new item is added to the catalog and no purely collaborative recommender can be applied.

In this paper, we show a practical application of current QCs technology for a Recommender System. We propose a formulation of a hybrid feature selection approach in a way that can be solved with a quantum annealer. We also report the results of experiments conducted using the D-Wave quantum annealer showing the proposed approach effectiveness and that current generation quantum annealers are now able to solve realistic problems.

The rest of the paper is organized as follows. In \secref{sec:related_works} we present related works for feature weighting and quantum approaches for recommender systems. In \secref{sec:recsys} we briefly introduce recommender systems and the recommendation problem. In \secref{sec:qa} we present the quantum computing technology we use.
In \secref{sec:cqfs} we present the proposed feature selection approach. In \secref{sec:eval} we discuss the results and scalability. Lastly, in \secref{sec:conclusions} we draw the conclusions.

\section{Related Works}
\label{sec:related_works}
Feature weighting methods, a generalization of feature selection, can be divided in three categories: \emph{filtering}, \emph{embedding} and \emph{wrapper} \cite{DBLP:journals/cee/ChandrashekarS14}.
Filter methods compute a score for a variable using a heuristic metric and then either use that score as weight or select only a certain number of the best weighted features. Many of such methods used in the recommender systems field originated from Information Retrieval, examples are the widely known TF-IDF and BM25. Other methods rely on the correlations between variables as well as on mutual information. These methods are lightweight but do not optimize a \mbox{predictive model.}

Embedding methods merge the feature weighting into the model training, learning both at the same time. Examples of this are Recursive Feature Elimination, which iteratively trains a model and then removes the least important features, but also Factorization Machines or Neural Networks which embed learning the feature weights as part of the model itself. An example of embedding method applied to the recommender systems field is {Feature-based factorized Bilinear Similarity Model} \cite{DBLP:conf/sdm/SharmaZHK15}, which learns to compute the similarities between items weighting the features according to how the users interacted with them. Since embedding methods optimize a certain predictive model, they can be more effective in selecting the important features tailored to the specific algorithm of interest, but this also makes them much more computationally expensive than filter methods.

Wrapper methods operate on a middle ground between the two, in which the feature weighting is not a part of the predictive model itself but its performance is used as part of an optimization problem to select the optimal weights for features. This optimization problem can be tackled with many tools, among which are heuristic greedy algorithms and genetic algorithms. In the recommender systems field examples of wrapper methods are {Collaborative-Filtering-enriched Content-Based Filtering (CFeCBF)} \cite{DBLP:journals/umuai/DeldjooDCECSIC19} and HP3 \cite{bernardis2018HP3}, which use a previously trained collaborative model and then learn feature weights in such a way that the content-based similarity would approximate the collaborative one. Another example of wrapper method, though not strictly a feature weighting one, used in a similar scenario is \emph{attribute to feature mapping} \cite{DBLP:conf/icdm/GantnerDFRS10}, which learns a mapping from features to the latent dimensions of a matrix factorization model.

There are only few articles applying quantum computing to recommender systems, most targeting the \emph{gate model} at a theoretical level. To the best of our knowledge, Ref. 
 \cite{DBLP:journals/corr/KerenidisP16} is the first article that proposes to use a quantum algorithm for low rank approximation models applied to recommender systems and claimed an exponential speedup over classical techniques. However, in a fruitful interaction between research fields, it was later shown that improved classical algorithms could close this speed gap to a polynomial one \cite{DBLP:conf/stoc/Tang19}. Another paper \cite{DBLP:journals/qip/ChakrabartyKS17} discusses the impact of Grover's algorithm, which is able to search in a non-structured database in sub-linear time, and its possible applications in recommender systems. Lastly, in \cite{DBLP:journals/amcs/SawerwainW19} it is proposed to combine a quantum-based KNN with the Grover algorithm.  
Other articles propose to use, instead, the \emph{adiabatic model}. For example, \mbox{Ref. \cite{DBLP:journals/fcomp/NishimuraTSMO19}} proposes a way to optimize the positioning of items in the page of a website both accounting for popularity and diversity, the proposed optimization problem is then solved on a \mbox{D-Wave QC.}

\section{Recommender Systems}
\label{sec:recsys}
The objective of a recommender system is to assist users in the exploration of the vast catalogs now at their disposal by recommending interesting items to users. It is known that using a well performing recommender systems improves user satisfaction.
The data on which they rely comprises of past user interactions with items, called collaborative information, and item attributes, called content information.
These kinds of data can be represented as matrices, on which recommendation models are built.
The {\emph{User Rating Matrix}} (URM) is a \twomatsize{U}{I} matrix containing user-item interactions.
These interactions can be explicit, with integer or real values corresponding to ratings, or implicit, with binary or categorical values representing a specific interaction with the items.
The {\emph{Item Content Matrix}} (ICM) is, instead, an \twomatsize{I}{F} matrix with binary values that indicate if an item has a certain feature, or real values if another kind of information is needed.
In this work we treated both explicit and implicit URMs, while we only used binary ICMs.

There are two main algorithm families used to build a recommendation model, collaborative filtering and content-based filtering.
They, respectively, use collaborative and content information. In this work we will consider in particular \emph{similarity-based} models:
\begin{equation*}
    \hat{R} = R \cdot S
\end{equation*}
where $R$ is the URM and $S$ an item--item similarity matrix computed with either collaborative or content information according to a certain recommendation model, \eg cosine similarity, linear regression. The product of $R$ and $S$ allows to compute the \emph{score} that the model attributes to each item given the user profile contained in $R$. The items are then ranked and the highest scored items are recommended to each user.

One of the possible ways to obtain the similarity model $S$ is by building an \emph{item-based k-nearest neighbor} model {\cite{sarwar2001item}}, which is the family of models we will focus on in this paper. 
Each value $S_{ij}$ can be computed as a cosine similarity between two vectors representing items $i$ and $j$:
\begin{equation*}
    S_{ij} = cos(\Vec{i}, \Vec{j}) = \frac{ \Vec{i} \cdot \Vec{j} }{ ||\Vec{i}||_2 \cdot ||\Vec{j}||_2 }
\end{equation*}
Then, only the $k$ highest similarities for each item are kept.
The model can also be tweaked with different hyperparameters.
For example, we could change the neighborhood size $k$, or change the similarity function.
This method is called \emph{ItemKNN} and is applicable both on collaborative and content information.
In the former case, item vectors contain interactions with the users, which are columns in the URM.
In the latter, item vectors correspond to rows in the ICM, containing their feature information.
Other methods are available for this task, such as matrix factorization methods or graph-based ones.

Notice that in a cold start scenario new items are added to the platform and therefore they have no collaborative information because no user has interacted with them yet. These items are called cold items and the only information available for them is the content one, contained in the ICM.

\section{Quantum Annealing}
\label{sec:qa}
The term \emph{quantum annealing} can be used to denote both a classical metaheuristic and its implementation as a physical device under the adiabatic QC paradigm.
Quantum annealing as a metaheuristic was first introduced by \citet{Apolloni:192546} in 1988, as a variant of simulated annealing \cite{DBLP:journals/ior/Pincus70, DBLP:journals/science/KirkpatrickGV83}.
Throughout this paper we will use quantum annealing to refer exclusively to the QC.

Quantum annealing exploits the natural tendency of every physical system to evolve towards and remain into its minimum energy state, thus, it can be used to find solutions for optimization problems that are represented in terms of the energy of a system with a certain structure.
In the Quantum Processing Unit (QPU) of a quantum annealer, the analogous of the CPU in a classical computer, qubits are connected to each other in a graph topology, establishing a physical system whose energy is measured by a function of its states, called \emph{Hamiltonian}. Further details on the physics and mathematical formulation can be found in \cite{PhysRevX.6.031015}.

\subsection{Quadratic Unconstrained Binary Optimization}
Quadratic unconstrained binary optimization (QUBO) is a class of problems that can be proved to be equivalent to the Hamiltonian of a quantum system.
A QUBO problem is defined as follows:
\begin{equation}
	\label{eq:qa:qubo}
	\begin{aligned}
		\min \quad & y = x^TQx \\
		& x \text{ binary}
	\end{aligned}
\end{equation}
where $ x \in \{0, 1\}^n $ is a vector of $ n $ binary variables and $ Q $ is an \matsize{n} matrix called QUBO matrix.
The QUBO formulation allows to describe NP-complete problems and some NP-hard optimization problems, for many of which there exists a simple formulation \cite{10.3389/fphy.2014.00005, DBLP:journals/4or/GloverKD19}. Since the QUBO formulation is equivalent to the Hamiltonian of a quantum system, any problem that can be represented as QUBO can be solved on a quantum annealer. {Note that, although quantum annealers are special-purpose devices that find low-energy solutions of Hamiltonians, in itself an NP-hard optimization problem, it is generally believed that due to their limitations when compared to other quantum computing paradigms they will not be able to lower the asymptotic complexity class of said problems.}

\subsection{The Annealer: D-Wave Advantage}
\begin{figure}[H]
	\includegraphics[width=0.475\linewidth]{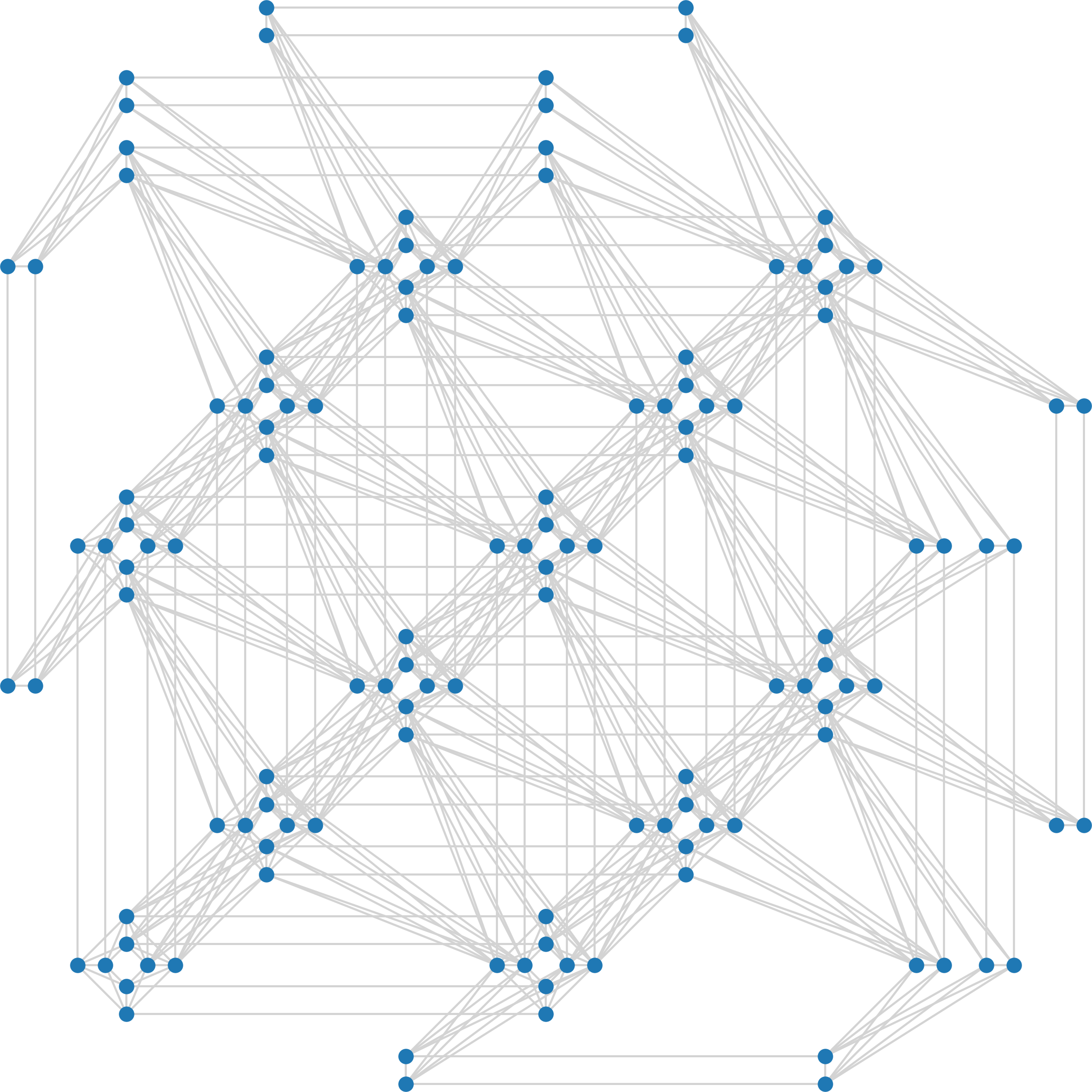}
	\caption{Portion of a Pegasus graph used in Advantage QPUs. Nodes represent qubits, while edges show the connections between qubits. The cross form of several Chimera cells is clearly visible.}
	\label{fig:qa:pegasus}
\end{figure}

\begin{figure}[H]
    \begin{tikzpicture}
        \node[shape=circle,draw=mplblue,text=mplblue] (0) at (0,3) {0};
        \node[shape=circle,draw=mplblue,text=mplblue] (1) at (0,2) {1};
        \node[shape=circle,draw=mplblue,text=mplblue] (2) at (0,1) {2};
        \node[shape=circle,draw=mplblue,text=mplblue] (3) at (0,0) {3};
        \node[shape=circle,draw=mplblue,text=mplblue] (4) at (2,3) {4};
        \node[shape=circle,draw=mplblue,text=mplblue] (5) at (2,2) {5};
        \node[shape=circle,draw=mplblue,text=mplblue] (6) at (2,1) {6};
        \node[shape=circle,draw=mplblue,text=mplblue] (7) at (2,0) {7};
        
        \path [-] (0) edge (4);
        \path [-] (0) edge (5);
        \path [-] (0) edge (6);
        \path [-] (0) edge (7);
        
        \path [-] (1) edge (4);
        \path [-] (1) edge (5);
        \path [-] (1) edge (6);
        \path [-] (1) edge (7);
        
        \path [-] (2) edge (4);
        \path [-] (2) edge (5);
        \path [-] (2) edge (6);
        \path [-] (2) edge (7);
        
        \path [-] (3) edge (4);
        \path [-] (3) edge (5);
        \path [-] (3) edge (6);
        \path [-] (3) edge (7);

        \node[shape=circle,draw=mplblue,text=mplblue] (0c) at (3,1.5) {0};
        \node[shape=circle,draw=mplblue,text=mplblue] (1c) at (3.75,1.5) {1};
        \node[shape=circle,draw=mplblue,text=mplblue] (2c) at (5.25,1.5) {2};
        \node[shape=circle,draw=mplblue,text=mplblue] (3c) at (6,1.5) {3};
        \node[shape=circle,draw=mplblue,text=mplblue] (4c) at (4.5,3) {4};
        \node[shape=circle,draw=mplblue,text=mplblue] (5c) at (4.5,2.25) {5};
        \node[shape=circle,draw=mplblue,text=mplblue] (6c) at (4.5,0.75) {6};
        \node[shape=circle,draw=mplblue,text=mplblue] (7c) at (4.5,0) {7};
        
        \path [-] (0c) edge (4c);
        \path [-] (0c) edge (5c);
        \path [-] (0c) edge (6c);
        \path [-] (0c) edge (7c);
        
        \path [-] (1c) edge (4c);
        \path [-] (1c) edge (5c);
        \path [-] (1c) edge (6c);
        \path [-] (1c) edge (7c);
        
        \path [-] (2c) edge (4c);
        \path [-] (2c) edge (5c);
        \path [-] (2c) edge (6c);
        \path [-] (2c) edge (7c);
        
        \path [-] (3c) edge (4c);
        \path [-] (3c) edge (5c);
        \path [-] (3c) edge (6c);
        \path [-] (3c) edge (7c);
    \end{tikzpicture}
    \caption{Chimera unit cell in its two common renderings, bipartite and cross form, with numbers representing qubits in the cell. We can see that each qubit is connected to 4 others of the same cell. Further connections, shown in \figref{fig:qa:chimera}, connect different cells.}
    \label{fig:qa:unit}
\end{figure}
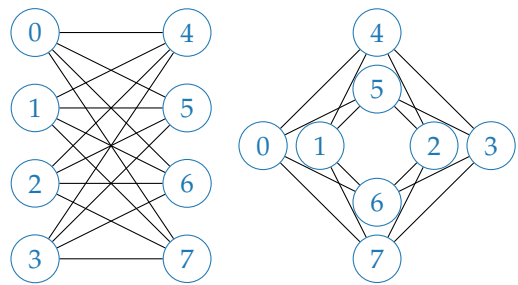

\begin{figure}[H]
	\includegraphics[width=0.5\linewidth]{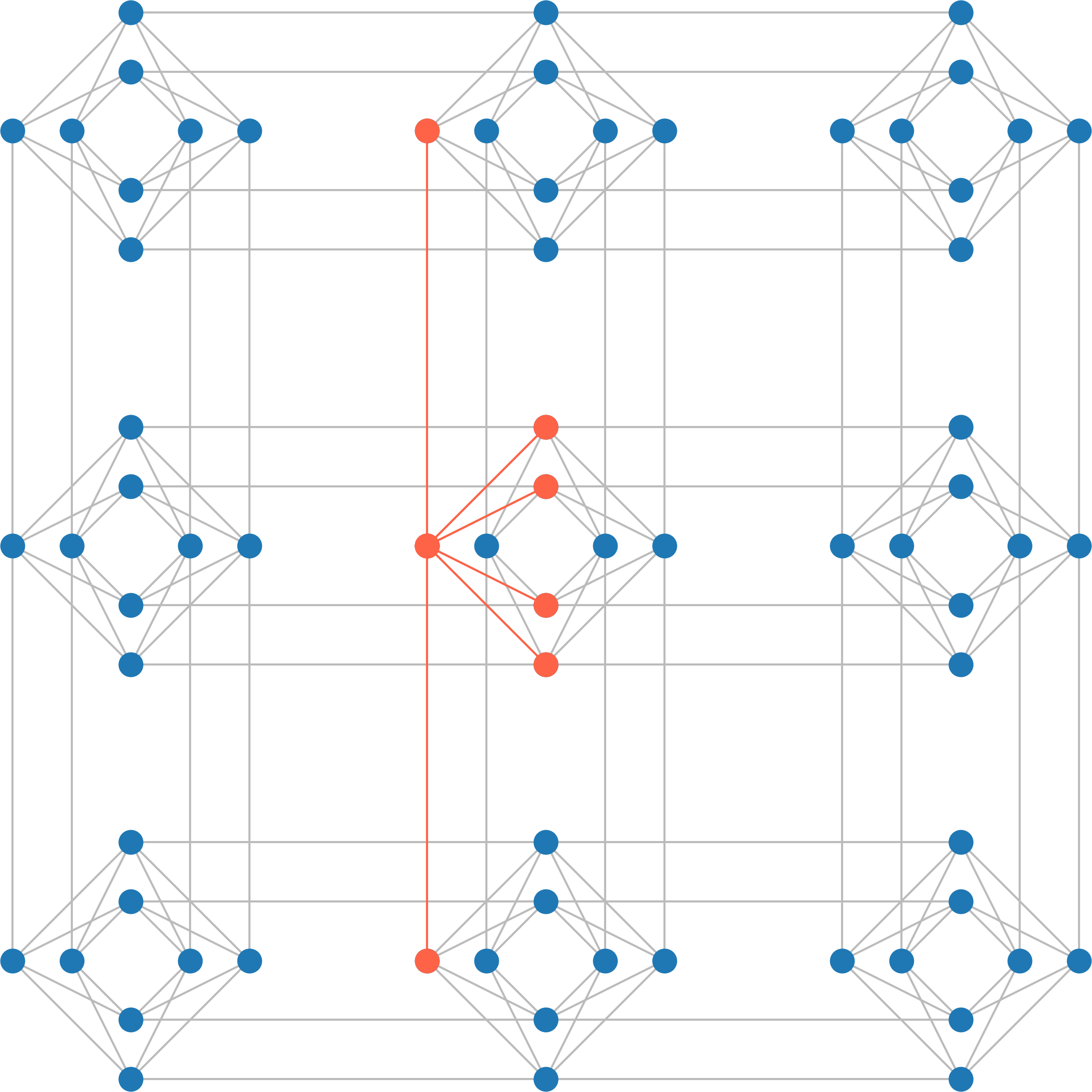}
	\caption{A Chimera graph of 9 unit cells, with a qubit and its connections highlighted, spanning both within a cell and between different cells.}
	\label{fig:qa:chimera}
\end{figure}

The physical annealer we used to carry out experiments for this work is the D-Wave Advantage, which offers more than 5000 qubits connected in a sparse graph called \mbox{Pegasus \cite{boothby2020next}}, shown in \figref{fig:qa:pegasus}. Each qubit is connected to a maximum of 15 others, some qubits in the external regions of the graph have less connections while others may have been deactivated due to defects or excessive noise.
The Pegasus topology is rather complex and a detailed description is beyond the scope of this paper. In order to guide the reader in understanding this new technology, we will describe a previous and simpler topology Pegasus is based on, \emph{Chimera}.
In the Chimera topology, qubits are laid out in bipartite graphs of eight nodes, called Chimera \emph{unit cells}, shown in \figref{fig:qa:unit}.
Each unit cell in a Chimera graph is connected to adjacent unit cells, so that each qubit is connected to six other qubits at most, as shown in \figref{fig:qa:chimera}. The Pegasus topology builds upon Chimera by adding further connections between qubits and cells, as can be seen in \figref{fig:qa:pegasus}.

\vspace{-12pt}

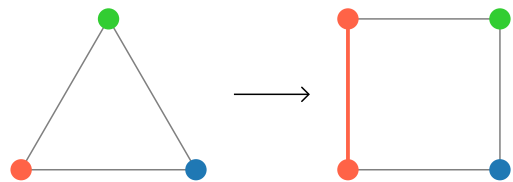
\begin{figure}[H]
    \begin{tikzpicture}
        
        \path [-, draw=gray, line width=0.2mm] (-0.3,0) edge (0.85,2);
        \path [-, draw=gray, line width=0.2mm] (0.85,2) edge (2,0);
        \path [-, draw=gray, line width=0.2mm] (2,0) edge (-0.3,0);

        \draw[fill=tomato, draw=none] (-0.3,0) circle (4pt);
        \draw[fill=limegreen, draw=none] (0.85,2) circle (4pt);
        \draw[fill=mplblue, draw=none] (2,0) circle (4pt);

        \path [-, draw=tomato, line width=0.5mm] (4,0) edge (4,2);
        \path [-, draw=gray, line width=0.2mm] (4,2) edge (6,2);
        \path [-, draw=gray, line width=0.2mm] (6,2) edge (6,0);
        \path [-, draw=gray, line width=0.2mm] (6,0) edge (4,0);

        \draw[fill=tomato, draw=none] (4,0) circle (4pt);
        \draw[fill=tomato, draw=none] (4,2) circle (4pt);
        \draw[fill=limegreen, draw=none] (6,2) circle (4pt);
        \draw[fill=mplblue, draw=none] (6,0) circle (4pt);
        
        \path[-angle 90, line width=0.2mm] (2.5,1) edge (3.5,1);
        
    \end{tikzpicture}
    \caption{Minor embedding of a triangular graph into a square topology. The triangular problem graph $G$ fits into the square topology $P$ by adding an edge and using two physical qubits to represent the same logical problem variable, here shown in red.}
    \label{fig:qa:embedding}
\end{figure}

In order to solve with the QPU a problem already represented in its QUBO formulation, we need to embed the QUBO problem into the QPU graph.
This can be done through a procedure called \emph{minor embedding} \cite{DBLP:journals/qip/Choi08}. 
The process has two main components: embedding and parameter setting.
Given a problem graph $G$, of which the QUBO matrix is the adjacency matrix, and the quantum annealer topology $P$, the goal of the embedding step is to find another graph $G_{emb}$, which is a subgraph of $P$ such that the original problem graph $G$ can be obtained from $G_{emb}$ by contracting edges.
An example of the embedding process is shown in \figref{fig:qa:embedding}, where we want to embed a triangular problem graph $G$ into a square topology $P$.
As we can see, the embedding algorithm generates a graph $G_{emb}$, which creates an edge and duplicates a node allowing the problem graph to fit the desired topology.
In this case, there will be two physical qubits representing the same logical variable. Clearly, we wish for the two physical qubits to have the same value.
The second step in the minor embedding process sets the strength of the connections between qubits to account for these additional nodes and edges created during the annealing process, \eg adds to the problem loss function new equality constraints between the qubits representing the same logical variable. 
It should be noted that the embedding is itself an NP-hard problem, however efficient heuristics that work in polynomial time have been specifically developed \cite{DBLP:journals/corr/CaiMR14}.

\vspace{-12pt}

\begin{figure}[H]
	\includegraphics[width=.8\linewidth]{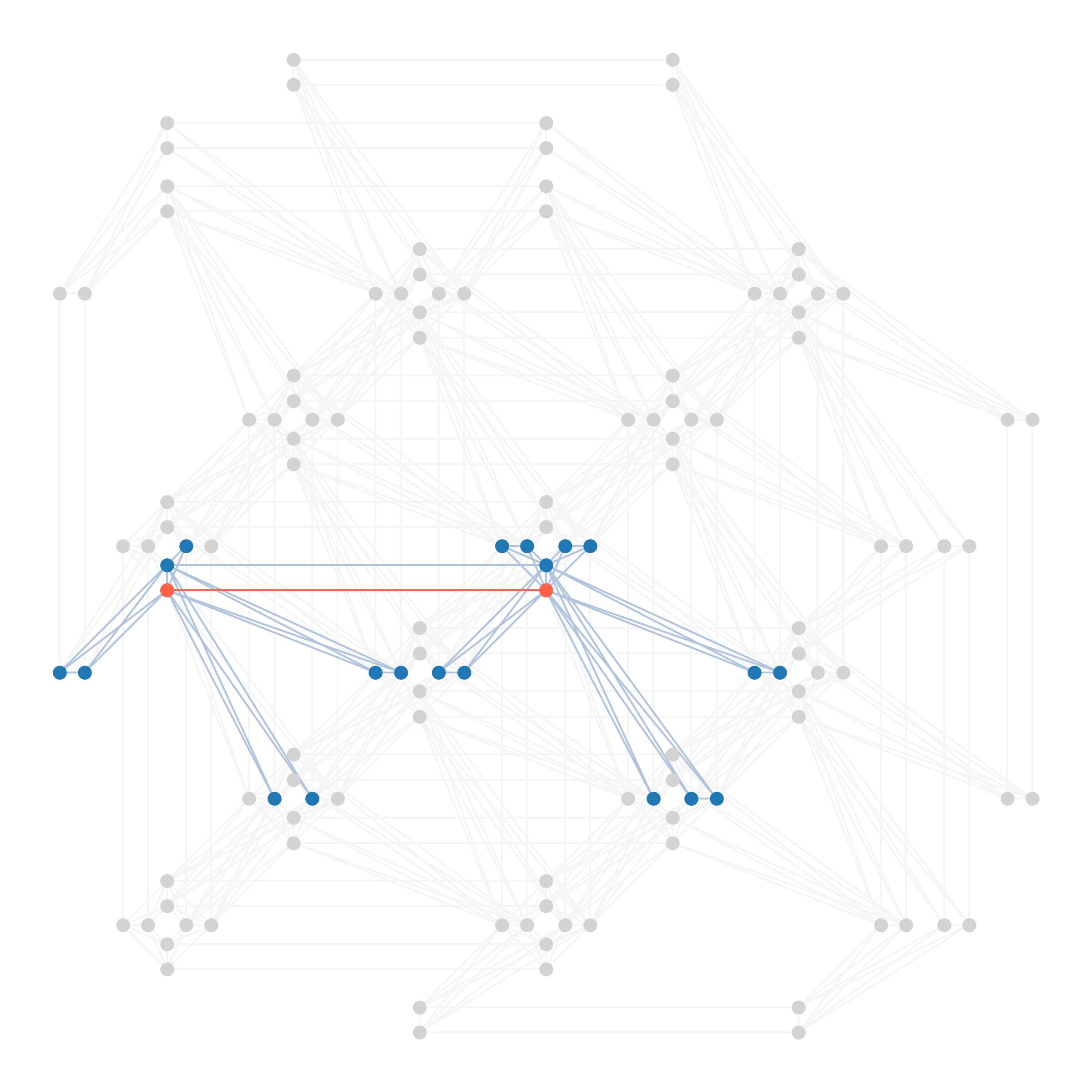}
	\caption{When a problem variable is connected to more variables than the topology allows, multiple physical qubits are used to represent a single logical variable in such a way that the connections can be split between them. In this example a problem variable is connected to 20 variables but the topology of the QPU only allows a maximum of 15. The problem variable is therefore represented using 2 physical qubits, referred to as a \emph{chain}, here represented in red.}
	\label{fig:qa:chain}
\end{figure}

The sparsity of the QPU graph may constitute a constraint if the problem that we wish to solve is denser than the QPU connectivity.
In such cases, additional auxiliary variables are automatically created during the minor embedding process, each using another qubit and lowering the number of free variables the problem can have. For example, since in the Pegasus topology each qubit is connected to 15 others, if we want to represent a constraint which connects a problem variable with 20 others then two different qubits will be needed to represent the same logical variable and the connections will be split between them, see \figref{fig:qa:chain}. As the connectivity of the problem increases so does the number of additional physical qubits that are needed to connect all the logical variables. The number of variables that can be embedded in a fully connected problem depends on both the graph topology, hence connectivity and size. In particular, for the current Pegasus topology with \mbox{5000 qubits}, the maximum number of variables that can be embedded in a fully connected problem is approximately 180. This value depends on the embedding algorithm, but can be estimated by using the formula $12M-10$, where M is a topological parameter, of value 16 for the current Pegasus topology.
We can look again to the Chimera topology for a simpler way to explain the topological parameter $M$, which was a function of the number of cells, that in Chimera were $M^2$. Different QPU generations may have different values for M.

An important feature of this technology is that the annealing time is very small and constant, in the range of $2\times10^{-5}$ s, allowing to find the solution of the desired problem very quickly. Due to the stochastic nature of the annealing process, multiple runs should be conducted, each producing a potentially different solution. The best solution can then be selected based on the corresponding loss function value or other criteria.
If the problem graph that we want to solve does not fit into the QPU, it is possible to combine the QPU with classical algorithms which can, for example, split the problem in smaller instances solvable by the QPU (dwave-hybrid---\url{https://docs.ocean.dwavesys.com/en/stable/docs_hybrid/sdk_index.html} (accessed on 27 July 2021)).
Hybrid solvers are also available through cloud services, but being a proprietary commercial solution, certain implementation details are not disclosed.

\section{Collaborative-Driven Quantum Feature Selection}
\label{sec:cqfs}
As explained in \secref{sec:qa}, quantum annealing can be used to solve NP-hard binary optimization problems with a QUBO formulation.
This formulation fits naturally the problem of feature selection \cite{10.1007/978-981-15-5616-6_10} (\url{https://github.com/dwave-examples/mutual-information-feature-selection} (accessed on 27 July 2021), \url{https://1qbit.com/whitepaper/optimal-feature-selection-in-credit-scoring-classification-using-quantum-annealer/} (accessed on 27 July 2021)).

We now describe our proposed approach \emph{Collaborative-driven Quantum Feature Selection} (CQFS).
The underlying idea of CQFS is that the features are not selected according to an information theoretic metric (\eg mutual information) or an information retrieval one (\eg TF-IDF) but rather, they should be selected according to how the users interact with them in order to leverage the domain knowledge which is implicit in their behavior.
In particular, we rely on a previously trained collaborative model and we perform feature selection in such a way that the content-based model built with the selected features mimics as closely as possible the collaborative one. Therefore, considering a content-based model built using all the features, if items that are similar in this model are similar also in the collaborative one, this is an indication that the features are important. On the contrary, if items are similar in the content-based model but are not in the collaborative one, this indicates that the features are not important and may add noise. A fundamental assumption of CQFS and of the other collaborative-based feature weighting methods CFeCBF and HP3 is that the quality of the collaborative model is superior to that of the content-based model for warm items, as it is often the case. The improved content-based model can then be used in scenarios where a collaborative model is not available, in particular in cold start situations.

First of all, we define the problem variables as $x_f \in \{0,1\}, f \in F$, where $F$ is the set of all item features, of cardinality $|F|$.
When $x_f = 1$, the corresponding feature $f$ is selected, while it is not if $x_f = 0$. 
From now on, we will refer to the vector of all binary variables corresponding to features as $x$. Furthermore, $I$ will be the set of all items and $|I|$ its cardinality.
Then, we define a QUBO problem as follows:
\begin{equation}
	\label{eq:cqfs:cqfs}
	\begin{aligned}
		\min \quad & y = x^T FPM x \\
		& x \text{ binary}
	\end{aligned}
\end{equation}
where $ FPM $ stands for \emph{Feature Penalization Matrix} and is an \fsize matrix containing the coefficients of the QUBO objective function.
The objective function can be obtained by directly comparing a collaborative and a content-based similarity model, respectively, represented by the matrices $S^{CF}$ and $S^{CBF}$, both of shape \isize.
For each couple of items $i, j \in I$, there are four distinct cases when comparing their content-based and collaborative similarities:

\begin{enumerate}
	\item Items $i$ and $j$ have both content-based and collaborative similarities greater than zero. The two similarities are therefore consistent, we want to keep the content similarity and therefore the common features of $i$ and $j$;
	\item Items $i$ and $j$ have a content-based similarity greater than zero but the collaborative one is zero. In this case we want to remove the features that generated the content similarity;
	\item Items $i$ and $j$ have a collaborative similarity greater than zero, but the content-based similarity is zero. Since the items have no common features we cannot change the content similarity;
	\item Items $i$ and $j$ have both content-based and collaborative similarities as zero. In this case the two similarities are already consistent.
\end{enumerate}

These four conditions establish the coefficients assigned to each couple of items when building the objective function.
The result of this comparison is represented with an auxiliary matrix called \emph{Item Penalization Matrix}, $ IPM $, of size \isize, which will later be used to represent the optimization problem in terms of features. \added{We can observe that the $ IPM $ could actually represent another QUBO problem and be used to select items. It should drive the selection towards items with non-zero similarity in both the collaborative and content-based models. At the same time, it should avoid the selection of items with a similarity present exclusively in the content-based model. Further discussion on this is however beyond the scope of the paper.}
Since this is a minimization problem, negative coefficients increase the probability that the corresponding problem variable is selected, wile positive coefficients reduce it.
Therefore, the coefficient $IPM_{ij}$ associated with the items $i$ and $j$, should be:
\begin{itemize}
	\item \textit{Positive}, if the similarity $S_{ij}$ between item $i$ and $j$ is present only in $S^{CBF}$;
	\item \textit{Negative}, if $S_{ij}$ is present both in $S^{CF}$ and in $S^{CBF}$;
	\item \textit{Zero} otherwise.
\end{itemize}

In \tabref{tab:ipm} we summarize the values that $IPM_{ij}$ should take depending on the four conditions previously described.

\begin{specialtable}[H]
	\caption{Value of the coefficient $IPM_{ij}$ for each possible condition on the collaborative and content-based similarities.}
	\label{tab:ipm}
	\setlength{\tabcolsep}{14.02mm}
	\begin{tabular}{cccc}
		\toprule
		& \boldmath{$S_{ij}^{CF}$} & \boldmath{$S_{ij}^{CBF}$} & \boldmath{$IPM_{ij}$} \\
		\midrule
		1. & >0 & >0 & <0 \\
		2. & =0 & >0 & >0 \\
		3. & >0 & =0 & $0$ \\
		4. & =0 & =0 & $0$ \\
		\bottomrule
	\end{tabular}
\end{specialtable}

In order to build the $IPM$, we treat positive and negative coefficients separately.
In particular, we call $K$ the \isize matrix containing the negative coefficients, associated with similarities we want to keep; we call $E$ the \isize matrix containing the positive coefficients, associated with the similarities we would like to eliminate from the model.
In particular, non-zero values in $K$ are equal to $-1$, while non-zero values in $E$ are equal to $1$.
Then, the $IPM$ is built as a weighted sum of these two matrices:
\begin{equation*}
	IPM = \alpha K + \beta E
\end{equation*}
where $\alpha$ and $\beta$ are the weighting parameters controlling how much the negative and positive components will influence the selection.

In order to perform feature selection, we now need to build the $ FPM $, used in the QUBO problem of \ceqref{eq:cqfs:cqfs}.
Similarities between items, in a content-based model, derive from the items' features.
Therefore, in order to preserve or remove a similarity, we need to encourage or discourage the selection of the two items' features, this should be reflected when building the $FPM$.

Let us consider two features $ f $ and $ g $, the corresponding problem variables $ x_f $ and $ x_g $ and the coefficient $ FPM_{fg} $ binding the two variables in the QUBO matrix.
Each couple of items $ i $ and $ j $, respectively, having features $ f $ and $ g $, should contribute to $ FPM_{fg} $ depending on the conditions met by $S_{ij}^{CF}$ and $S_{ij}^{CBF}$.
In particular, the contribution derived from $i$ and $j$ is already represented in the $ IPM $.
Therefore, the value for $ FPM_{fg} $ should correspond to the summation of the values in the $IPM$ for all couples of items $i$ and $j$ that, respectively, have features $f$ and $g$.
Let $ICM \in \{0, 1\}^{|I|\times|F|} $ be the item content matrix. Therefore, $ FPM_{fg} $ can be computed as:
\begin{equation*}
	\label{eq:cqfs:qfg}
	\begin{aligned}
		FPM_{fg} &= \sum_{i \in I} \sum_{j \in I} ICM_{if} \cdot ICM_{jg} \cdot IPM_{ij} \\
		&= \sum_{i \in I} ICM_{if} \sum_{j \in I} IPM_{ij} \cdot ICM_{jg}
	\end{aligned}
\end{equation*}
It is straightforward to translate  the last equation into a matrix notation:
\begin{equation*}
	\label{eq:cqfs:fpm}
	FPM = ICM^T \cdot IPM \cdot ICM
\end{equation*}

As an additional step in our model, we impose a soft constraint into the objective function in order to select a specific percentage $p \in [0,1]$ of the original number of features. The added penalty is the following:
\begin{equation*}
	\label{eq:cqfs:kcomb}
	c = s \cdot \left(\sum_{f = 1}^{|F|} x_f - k\right)^2
\end{equation*}
where $ s $ is a parameter called \textit{strength}, controlling the magnitude of the penalty, and $ k = p |F| $.
This penalty is minimized for each of the $k$-combinations of its variables, \idest when the desired number of features is selected.
The resulting formulation of CQFS is:
\begin{equation}
	\begin{aligned}
		\min \quad & y = x^T FPM x + c \\
		& x \text{ binary}
	\end{aligned}
\end{equation}
which considers both the $ FPM $ and the $k$-combinations constraints.
When solved, this binary optimization problem gives a selection of features, which is then used to train a content-based recommendation model.

\section{Evaluation and Results}
\label{sec:eval}
In order to evaluate the proposed approach we test it on a known dataset with both collaborative and content feature information. In this section, we report the results of a content-based recommender system built on features selected by CQFS, both in terms of accuracy and beyond-accuracy metrics. We also discuss the computational time and scalability of the proposed approach and of the quantum annealer. We release the source code of our experiments in a public repository (\url{https://github.com/qcpolimi/CQFS} (accessed on 27 July 2021)). 
Notice that the quantum and hybrid solvers used in the experiments are freely available through D-Wave cloud services (\url{https://www.dwavesys.com/take-leap} (accessed on 27 July 2021)).

\subsection{Datasets}
Three datasets with features were used to carry out experiments in this work, \emph{The Movies Dataset}, \emph{CiteULike-a} and \emph{Xing Challenge 2017}.

The Movies Dataset is a dataset publicly available on Kaggle (The Movies Dataset---\url{https://www.kaggle.com/rounakbanik/the-movies-dataset} (accessed on 27 July 2021)).
It consists of an ensemble of data coming from a movies information database TMDb (The Movie Database (TMDb)---\url{https://www.themoviedb.org/} (accessed on 27 July 2021)) and the Full MovieLens Dataset by GroupLens.
It contains user ratings and various features of movies released up to July 2017.
In particular, in this work we use \replaced{the following metadata features: genre, original language, production company, production country, release date, spoken language, movie collection, adults movie (or not), status (rumored/planned/in production/post production/released), with video (or not)}{metadata features (The features we used are: genre, original language, production company, production country, release date, spoken language, movie collection, adults movie (or not), status (rumored/planned/in production/post production/released), with video (or not))}.
The dataset is preprocessed to include only features appearing in at least 5 items.
After preprocessing the dataset contains 270,882 user, 44,711 items and \mbox{3058 features.}

CiteULike-a is a dataset coming from the site CiteULike, a reference manager and bibliography sharing service for scientific papers.
It is presented with this name in \cite{DBLP:conf/kdd/LiS17} and contains information about users and articles.
Collaborative information consists of articles bookmarked by the users.
Content information associates articles with tags extracted from their titles and abstracts.
The dataset contains 5551 users, 16,980 items and 8000 features.

Xing Challenge 2017 is the dataset used in the 2017 ACM RecSys Challenge (ACM RecSys Challenge 2017---\url{http://www.recsyschallenge.com/2017/} (accessed on 27 July 2021)), released by Xing, a business social network.
It contains interactions between the platform's users and job postings from companies, which in this recommendation scenario are the items.
Interaction data is preprocessed, in order to keep only users and items having at least five interactions.
Of all the features present in the dataset, we use features indicating industry, discipline, career level, employment type, country and region of a job posting.
After preprocessing the dataset contains 190,291 users, 88,984 items and 79 features.

\subsection{Experimental Pipeline}
The proposed feature selection algorithm is meant to operate in a cold start setting, hence for items that do not have collaborative information. However, it must be trained in a warm start setting, where collaborative information for items is available. This means that it requires a two-step pipeline in which first the cold items are removed from the training data and then, with the warm items, the optimization of the collaborative models and the selection of the features can take place. In this section, we describe the various steps in the experimental pipeline.

\subsubsection{Data Splitting}
As previously described, the data requires first a cold item split and then a warm item split for the collaborative models.

\begin{figure}[H]
\scalebox{0.6}{
\begin{tikzpicture}[every node/.style={minimum size=.3cm-\pgflinewidth, outer sep=0pt}]
	\node[scale=1.5] at (3,-0.5) {warm items};
    \draw[draw=black, dashed, ultra thick] (0,0) -- (0,-1);
    \draw[draw=black, dashed, ultra thick] (0,-1) -- (6,-1);
    \draw[draw=black, dashed, ultra thick] (6,-1) -- (6,0);
    
	\node[scale=1.5, rotate=90] at (-0.5,2) {users};
    
    \draw[draw=black, ultra thick] (0,0) rectangle (6,4);
    \draw[draw=black, ultra thick] (6,0) rectangle (8,4);
    \draw[draw=black, ultra thick] (8,0) rectangle (10,4);
    
    \node[minimum size=1.cm] at (3,2) {\LARGE Training};
    \node[minimum size=1.cm] at (7,2) {\LARGE Valid.};
    \node[minimum size=1.cm] at (9,2) {\LARGE Test};
\end{tikzpicture}
}\caption{Data splitting done item-wise, the training and validation split contains the warm items while the test set contains the cold items.}
\label{fig:cold_item_split}
\end{figure}
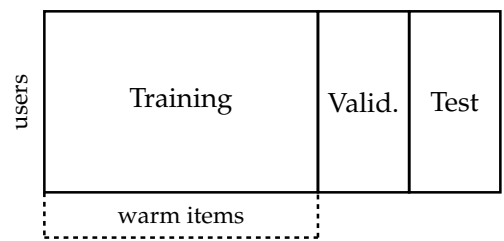

\begin{description}
    \item[Cold item split:] is used to obtain a cold test and validation set for the hyperparameter optimization and testing of the feature weighting or selection algorithms. We iteratively select a random item until the set of selected items accounts for a certain percentage of the original dataset interactions, see \figref{fig:cold_item_split}. The cold test data accounts for 20\% of the interactions, the validation for 10\%. The union of this training and validation data is considered as warm item set.
	\item[Collaborative data split:] is used for the training and optimization of the collaborative models. The previous warm item set is split by randomly sampling 10\% of the interactions in each user profile for the collaborative validation data. \added{Note that it could be possible to apply this random holdout split solely on the previous cold item training data. We chose to include the validation data to avoid the collaborative training data becoming too small.}
\end{description}

\subsubsection{Collaborative Models}
Being CQFS a wrapper method, it requires a similarity matrix generated with another model. 
The collaborative algorithms are trained on the collaborative training data and their hyperparameters optimized on the collaborative validation data.
We selected a set of simple but effective collaborative models from different families of methods \cite{DBLP:conf/recsys/DacremaCJ19}:
\begin{description}
    \item[ItemKNN:] the widely known item-based nearest-neighbor method using cosine similarity and shrinkage to reduce the similarities of items with low support \cite{sarwar2001item}. Its hyperparameters are the shrink term, the number of neighbors and whether to use a simple weighting strategy for the items.
    \item[PureSVD:] simple matrix factorization model proposed in \cite{DBLP:conf/recsys/CremonesiKT10} that decomposes the user-item interaction data in the product of two lower dimensionality matrices using the standard SVD decomposition. Via folding-in, \idest the product of the item latent factors by its transpose, PureSVD is equivalent to an item-item similarity model. Its only hyperparameter is the number of latent factors.
    \item[\rptb:] a graph-based collaborative similarity model based on a bipartite graph of users and items proposed in \cite{DBLP:journals/tiis/PaudelCNB17}. \rptb builds an item-item similarity model which represents the transition probability between the two items by simulating a random walk, this similarity is elevated to a power $\alpha$ and divided by the popularity of the two items elevated to a power $\beta$, which controls the popularity bias. Its hyperparameters are the coefficients $\alpha$, $\beta$ and the neighborhood size.
\end{description}

The hyperparameters of the collaborative models have been optimized with Bayesian search, by exploring 50 cases with the first 15 used as random initialization and optimizing Precision at cutoff 10. The ranges and distributions of the hyperparameters are those reported in \cite{10.1145/3434185}, see \tabref{tab:eval_hyperparams}.

\subsubsection{Quantum Feature Selection}
Given the collaborative similarity, the CQFS minimization problem is built as described in \secref{sec:cqfs}.
The model has the following hyperparameters: $\alpha$, weight of the similarities (hence features) to keep component, $\beta$, associated to the similarities (hence features) to remove component, $p$ the percentage of features we select and $s$ the strength for the constraint on the number of features. The range and distribution of those hyperparameters is reported in \tabref{tab:eval_hyperparams}.
Notice that parameter $p$ for Xing Challenge 2017 does not take values 20\% and 30\% because of the already small number of features in the dataset.
This optimization problem formulated as a QUBO can be solved with quantum annealing, classical techniques and hybrid quantum-classical ones. The hyperparameters of the QUBO models are optimized via a grid search. The resulting selection is then used to train an ItemKNN content-based model. The hyperparameters resulting in the best recommendation quality for the final ItemKNN content-based model evaluated on the cold validation split are selected. The quantum annealer can generate many different solutions in a very short time due to its stochastic nature. We sampled 100 solutions and selected the one with the best fitness value according to the defined QUBO model.
This device also offers various hyperparameters related to the underlying physical system, \eg the duration of the evolution, its schedule (the details of the predefined evolution schedule can be found here: \url{https://support.dwavesys.com/hc/en-us/articles/360005267253-What-are-the-Anneal-Schedule-Details-for-this-QPU} (accessed on 27 July 2021)) {and time-offsets for each qubits. The impact of this hyperparameters on the solution quality is still not well-understood and is an open research question \cite{PhysRevApplied.11.044083}; therefore, it is outside the scope of this work. In our experiments we used the default hyperparameters and settings for the quantum annealer (\idest annealing duration 20 $\upmu$s and we did not set a custom annealing schedule).}
The hybrid solver instead samples only one solution, which corresponds to the best loss obtained. We did not run the hybrid solver multiple times due to resource constraints. Similarly, other classical approaches, \idest TF-IDF and CFeCBF only produce one solution by design.

\subsubsection{Final Content-Based Model}

We use the result of the feature selection phase to train and optimize an ItemKNN content-based model using cosine similarity and shrinkage. In this case as well the hyperparameters are optimized  with 50 cases of Bayesian search following the indications and hyperparameter ranges reported in \cite{10.1145/3434185}.
The hyperparameters are optimized by training the model on the cold training split and selecting the hyperparameters optimizing the recommendation quality on the cold validation data. Once the optimal hyperparameters have been selected, the content-based model is trained again on the union of training and validation data and the results evaluating it on the test data are reported.

\subsection{Baselines}
We compare the CQFS results with three baseline algorithms:
\begin{description}
    \item[ItemKNN CBF:] item-based nearest-neighbor model using cosine similarity on all the original item features and shrinkage. Its hyperparameters are the same of the collaborative ItemKNN. In this case, weighting strategies such as TF-IDF and BM25 can be applied to the features. \vspace{-5pt}
    \item[TF-IDF:] a variant of ItemKNN CBF in which the features are selected as a certain quota (40\%, 60\%, 80\%, 95\%) of those with the highest TF-IDF score, a widely known feature weighting approach. Since TF-IDF is a filtering method it is fast to compute but does not optimize a predictive model. It has the same hyperparameters as ItemKNN CBF. 
    \vspace{-5pt}
    \item[CFeCBF:] a hybrid content-collaborative feature weighting algorithm proposed in \cite{DBLP:journals/umuai/DeldjooDCECSIC19}. CFeCBF shares similarities with our proposed approach in that it learns feature weights to approximate a collaborative model. The model has several hyperparameters related to the selection of which similarity values to learn, the gradient descent as well as the regularization and neighborhood size similarly to ItemKNN CBF. The optimal number of epochs has been selected via early-stopping by optimizing accuracy on the validation data, evaluating every five epochs and terminating after five consecutive evaluations in which the best results did not improve. The detailed hyperparameters are reported in \tabref{tab:eval_hyperparams}.
\end{description}

In order to ensure a fair comparison, all baselines are optimized with the same data splits used for CQFS, using the same Bayesian search with 50 cases as done for the collaborative models.

\begin{specialtable}[H]
    \tablesize{\small}
        \caption{Hyperparameter ranges used for the optimization of all the reported algorithms, both baselines and CQFS.}
        \label{tab:eval_hyperparams}
        \setlength{\tabcolsep}{4.26mm}
        \begin{tabular}{cccc}
            \toprule
            \textbf{Algorithm}	& \textbf{Hyperparameter}	&  \textbf{Range}	 & \textbf{Distribution}	\\ 
            \midrule
            \multirow{4}{*}{\begin{tabular}{c}ItemKNN \\ cosine\end{tabular}}  	
            				&topK	        & 5--1000 	& uniform 	\\ 
            				&shrink	        & 0--1000 	& uniform 	\\ 
            				&normalize \emph{$^{a}$} 	    & True, False 	&  categorical	\\ 
            				&weighting	& none, TF-IDF, BM25 	& categorical	\\ 
            \midrule
            \multirow{1}{*}{PureSVD}  	
            				&num factors	& 1--350 	& uniform 	\\ 
            \midrule
            \multirow{4}{*}{\rptb}  	
            				&topK	        & 5--1000 	& uniform 	\\ 
            				&alpha	        & 0--2	& uniform 	\\ 
            				&beta	        & 0--2	& uniform 	\\ 
            		        &normalize \emph{$^{b}$} 	& True, False 		&  categorical	\\ 
        
            \midrule
            \multirow{9}{*}{CFeCBF}  	
                            &epochs	        & 1--300                 	& early-stopping 	\\
                            &learning rate  & $10^{-5}$--$10^{-2}$    & log-uniform 	\\ 
                            &sgd mode	    & Adam                     &  categorical	\\
                            &$l1$ reg   	& $10^{-2}$--$10^{+3}$   & log-uniform 	\\ 
                            &$l2$ reg   	& $10^{-1}$--$10^{+3}$  & log-uniform 	\\     
                            &dropout        & 30--80\%            & uniform 	\\   
                            &initial weight & 1.0---random      	&  categorical	\\
                            &positive only  & True, False 	           	&  categorical	\\ 
                            &add zero quota \emph{$^{c}$} & 50--100\%             & uniform 	\\   
            \midrule
            
            \multirow{4}{*}{CQFS}  	
                    		&$ \alpha $  & $ 1 $   & categorical\\
                    		&$ \beta $   & $ 10^{0} $--$ 10^{-4}$  & log-uniform \\
                    		&$ s $       & $ 10^{0}$--$ 10^{+4} $ & log-uniform  \\
                    		&$ p $       & $ 40\%, 60\%, 80\%, 95\% $  & categorical \\
        	\bottomrule
       	\end{tabular}
\pbox{\columnwidth}{\footnotesize \emph{$^{a}$} \label{foot:knn_normalize}The \emph{normalize} hyperparameter in KNNs controls whether to use the denominator of the cosine similarity. If False, the similarity becomes the dot product alone. \emph{$^{b}$} \label{foot:normalize_similarity}The \emph{normalize} hyperparameter in \rptb refers to applying $l1$ regularization on the rows of the similarity matrix after the selection of the neighbors, to ensure it still represents a probability distribution. \emph{$^{c}$} Percentage of item similarities of value zero added as negative samples to improve its ranking performance.}
\end{specialtable}

\subsection{Results}
In this section, we present the results obtained with an ItemKNN CBF trained using the features selected by CQFS on The Movies Dataset, Xing Challenge 2017 and CiteULike-a.
Experiments were performed both with quantum-based and classical solvers for QUBO problems.
In particular, the quantum-based solver used for The Movies Dataset and CiteULike-a is D-Wave’s \emph{Leap Hybrid V2} service. For Xing Challenge 2017 we used the D-Wave \emph{Advantage} quantum annealer, since the QUBO problem in this case is completely embeddable onto the QPU.
The chosen classical solver, instead, is Simulated Annealing (SA) (dwave-neal---\url{https://docs.ocean.dwavesys.com/en/stable/docs_neal/} (accessed on 27 July 2021)).
We evaluate all models with traditional accuracy and ranking metrics: Precision, Recall, NDCG, MAP as well as beyond-accuracy metrics: Item Coverage, Gini Diversity and Mean Inter List diversity (MIL) \replaced{\cite{zhou2010solving, DBLP:conf/aaai/Dacrema21}}{\cite{zhou2010solving}} and report the results at a list length of 10.

First, we verify that the collaborative models provide higher recommendation quality than the content-based one on the warm items. With respect to The Movies Dataset, if we compare the recommendation quality on the collaborative validation data we can see that the ItemKNN CBF model has and NDCG@10 of 0.026, while the collaborative models have an NDCG@10 between 0.169 and 0.195. For CiteULike-a, the collaborative accuracy is only doubled, compared to the content-based accuracy, while for Xing Challenge 2017 the collaborative model outperforms the content one by three orders of magnitude. We also compare the structure of the content-based and collaborative similarity matrices. On The Movies Dataset we can see that, on average, only 0.23\% of couples of items have both a content and a collaborative similarity, 65.4\% have only a content similarity, a negligible amount has only a collaborative similarity and, lastly, 34.3\% have zero similarity in both. Xing Challenge 2017 presents very close percentages to these, while in CiteULike-a around 2\% of the item couples have both a content and a collaborative similarity, 52.7\% only content and 45\% none.

The results for The Movies Dataset are reported in \tabref{tab:eval:results:tmd:hv2}.
We can see that both the TF-IDF and CFeCBF baselines do not improve over the simple ItemKNN CBF algorithm, in this setting. In some cases, CFeCBF is more accurate than TF-IDF, which denotes the importance of choosing which collaborative model to use in the algorithm. On the other hand, CQFS is able to outperform the baselines in different settings when trained on the ItemKNN and PureSVD collaborative models. When training using \rptb, CQFS is not able to outperform the baseline ItemKNN CBF algorithm but its recommendation quality for $p=60\%$ is almost identical when measured with MAP and within 1.7\% when measured with other accuracy metrics. We can also observe how for the versions trained using ItemKNN and PureSVD there exist several solutions outperforming the baselines at different quotas of selected features and with hyperparameters that sometimes differ by an order of magnitude. This indicates that CQFS is rather robust to the selected hyperparameters. Regarding its ability to select a limited number of important features, we can see how CQFS is able to outperform or remain comparable with the ItemKNN baseline while selecting only 60\% of features when using the ItemKNN as collaborative model and even as little as 30\% of features when using the PureSVD model. When using the PureSVD collaborative model it is possible to select as low as 20\% of features losing only 2\% of NDCG. This indicates that the proposed approach is effective in identifying a few of highly important features without affecting the model quality.
Interestingly, despite removing the majority of features, this does not negatively affect the diversity of the recommendations. Despite being seldom reported in evaluation, beyond-accuracy metrics are important to measure how the recommendation model is diversifying its recommendations. Since the purpose of a recommender system is to help the user explore a catalog in a personalized way, ensuring the model generates diverse recommendations is important.  
On the other hand, CFeCBF results in lower recommendation quality as well as lower diversity, suggesting the model is more prone to popularity bias \added{(note that the paper proposing CFeCBF evaluated it too on a variant of MovieLens as dataset but relied on multimedia features)}.
The ability of CQFS to discard the majority of features without affecting the recommender effectiveness, both in terms of accuracy and in terms of diversity, could also be beneficial in terms of the scalability of the recommender algorithm itself.

{Although QCs could be in principle able to find better optimal solutions compared to classical methods, today's QCs are limited by noise, which can degrade the solution quality. QCs of the gate model are particularly vulnerable to it as the information encoded in qubits is rapidly lost, in a process known as decoherence. Quantum annealers, on the other hand, are more robust to noise due to the different physical process they leverage.} {In order to assess whether the noise is negatively affecting the solution quality, we compare the solutions found by the QPU with a classical Simulated Annealing solver, see \mbox{\tabref{tab:eval:results:tmd:sa}.}} 
 The results refer to the same setting of \tabref{tab:eval:results:tmd:hv2}, hence the baselines are not included as they would be identical. We can see that the results obtained with Simulated Annealing are in line with those obtained by the hybrid QPU solver, indicating that the QPU has been a reliable solver in this setting. {There is not yet conclusive evidence that the QPU provides better solutions than Simulated Annealing but this is nonetheless an important indication that this rapidly developing technology is now able to compete with classical solvers on realistic problems. It is expected that it will become more likely to observe improved solution quality using the QPU, compared to classical techniques, as the technology improves. It is however already possible to see promise in terms of its scalability, see Section \ref{sec:scalability}.}

\end{paracol}
\nointerlineskip
\begin{specialtable}[H]
    \small
    \widetable
	\caption{Results on The Movies Dataset. ItemKNN CBF, TF-IDF and CFeCBF are baselines, CQFS the proposed method. Both TF-IDF and CQFS refer to an ItemKNN CBF model trained on the features selected by the corresponding method. Recommendation lists have length 10.The last three columns contain CQFS best hyperparameters chosen with respect to recommendation quality on the validation set. CQFS metrics outperforming baselines are highlighted in bold.}
	\label{tab:eval:results:tmd:hv2}
	\setlength{\tabcolsep}{3.52mm}
	\begin{tabular}{ccccccccccc}
		\toprule
		\textbf{Models} & \textbf{Precision} & \textbf{Recall} & \textbf{NDCG} & \textbf{MAP} & \textbf{I. Cov.} & \textbf{Gini} & \textbf{MIL} & \boldmath{$\alpha$} & \boldmath{$\beta$} & \boldmath{$s$} \\
		\midrule
		ItemKNN CBF & 0.1218 & 0.0795 & 0.0856 & 0.0808 & 0.6637 & 0.0578 & 0.9566 & - & - & - \\
		\midrule
		TFIDF 40\% & 0.0503 & 0.0177 & 0.0269 & 0.0337 & 0.1628 & 0.0146 & 0.8347 & - & - & - \\
		TFIDF 60\% & 0.0793 & 0.0424 & 0.0595 & 0.0541 & 0.2251 & 0.0222 & 0.8923 & - & - & - \\
		TFIDF 80\% & 0.0838 & 0.0477 & 0.0633 & 0.0565 & 0.3186 & 0.0381 & 0.9411 & - & - & - \\
		TFIDF 95\% & 0.0914 & 0.0508 & 0.0665 & 0.0634 & 0.4510 & 0.0505 & 0.9436 & - & - & - \\
		\midrule
		CFeCBF ItemKNN & 0.0893 & 0.0574 & 0.0692 & 0.0598 & 0.5962 & 0.0451 & 0.9344 & - & - & - \\
		CFeCBF PureSVD & 0.0818 & 0.0547 & 0.0681 & 0.0577 & 0.4704 & 0.0309 & 0.9135 & - & - & - \\
		CFeCBF \rptb & 0.0751 & 0.0481 & 0.0563 & 0.0499 & 0.6325 & 0.0472 & 0.9429 & - & - & - \\
		\midrule
		\multicolumn{11}{c}{CQFS trained with ItemKNN} \\
		\midrule
		CQFS 20\% & 0.1124 & 0.0748 & 0.0788 & 0.0725 & 0.6552 & 0.0548 & 0.9545 & 1 & $10^{-4}$ & $10^2$ \\
		CQFS 30\% & 0.1163 & 0.0767 & 0.0819 & 0.0767 & 0.6558 & 0.0552 & 0.9552 & 1 & $10^{-4}$ & $10^3$ \\
		CQFS 40\% & 0.1208 & 0.0793 & 0.0853 & 0.0806 & 0.6486 & 0.0553 & 0.9555 & 1 & $10^{-3}$ & $10^2$ \\
		CQFS 60\% & {\textbf{0.1233}} 
 & {\textbf{0.0802}} & {\textbf{0.0867}} & {\textbf{0.0828}} & 0.6531 & 0.0563 & 0.9571 & 1 & $10^{-4}$ & $10^2$ \\
		CQFS 80\% & \textbf{0.1232} & \textbf{0.0802} & \textbf{0.0865} & \textbf{0.0824} & 0.6564 & 0.0571 & 0.9576 & 1 & $10^{-3}$ & $10^2$ \\
		CQFS 95\% & \textbf{0.1225} & \textbf{0.0798} & \textbf{0.0861} & \textbf{0.0814} & 0.6631 & 0.0577 & 0.9568 & 1 & $10^{-3}$ & $10^3$ \\
		\midrule
		\multicolumn{11}{c}{CQFS trained with PureSVD} \\
		\midrule
		CQFS 20\% & 0.1191 & 0.0779 & 0.0838 & 0.0787 & 0.6359 & 0.0548 & 0.9554 & 1 & $10^{-4}$ & $10^2$ \\
		CQFS 30\% & \textbf{0.1223} & \textbf{0.0798} & \textbf{0.0859} & \textbf{0.0815} & 0.6366 & 0.0555 & 0.9563 & 1 & $10^{-4}$ & $10^2$ \\
		CQFS 40\% & \textbf{0.1219} & 0.0795 & \textbf{0.0857} & \textbf{0.0811} & 0.6378 & 0.0556 & 0.9562 & 1 & $10^{-4}$ & $10^3$ \\
		CQFS 60\% & \textbf{0.1232} & \textbf{0.0801} & \textbf{0.0861} & \textbf{0.0820} & 0.6506 & 0.0568 & 0.9577 & 1 & $10^{-4}$ & $10^3$ \\
		CQFS 80\% & 0.1199 & 0.0783 & 0.0841 & 0.0789 & 0.6609 & 0.0577 & 0.9568 & 1 & $10^{-3}$ & $10^3$ \\
		CQFS 95\% & \textbf{0.1228} & \textbf{0.0801} & \textbf{0.0863} & \textbf{0.0818} & 0.6617 & 0.0577 & 0.9571 & 1 & $10^{-3}$ & $10^2$ \\
		\midrule
		\multicolumn{11}{c}{CQFS trained with \rptb} \\
		\midrule
		CQFS 20\% & 0.1085 & 0.0712 & 0.0702 & 0.0677 & 0.6951 & 0.0594 & 0.9598 & 1 & $10^{-3}$ & $10^2$ \\
		CQFS 30\% & 0.1097 & 0.0710 & 0.0710 & 0.0689 & 0.6999 & 0.0605 & 0.9607 & 1 & $10^{-4}$ & $10^2$ \\
		CQFS 40\% & 0.1111 & 0.0720 & 0.0730 & 0.0706 & 0.7011 & 0.0612 & 0.9609 & 1 & $10^{-4}$ & $10^3$ \\
		CQFS 60\% & 0.1198 & 0.0781 & 0.0848 & 0.0803 & 0.7056 & 0.0609 & 0.9583 & 1 & $10^{-4}$ & $10^2$ \\
		CQFS 80\% & 0.1193 & 0.0777 & 0.0844 & 0.0795 & 0.7101 & 0.0617 & 0.9587 & 1 & $10^{-4}$ & $10^3$ \\
		CQFS 95\% & 0.1200 & 0.0785 & 0.0842 & 0.0790 & 0.6684 & 0.0578 & 0.9568 & 1 & $10^{-3}$ & $10^2$ \\
		\bottomrule
	\end{tabular}
\end{specialtable}
\begin{paracol}{2}
\switchcolumn
\vspace{-12pt}

\begin{specialtable}[H]
	\caption{A selection of CQFS results {on The Movies Dataset} when the QUBO problem is solved with Simulated Annealing, showing the obtained results are comparable with the Hybrid QPU solver.}
	\small
	\label{tab:eval:results:tmd:sa}
	\setlength{\tabcolsep}{4.61mm}
	\begin{tabular}{ccccccc}
		\toprule
		\textbf{Models} & \textbf{Precision} & \textbf{Recall} & \textbf{NDCG} & \textbf{MAP} & \textbf{I. Cov.} & \textbf{MIL} \\
		\midrule
		\multicolumn{7}{c}{CQFS trained with PureSVD} \\
		\midrule
		CQFS 20\% & 0.1170 & 0.0764 & 0.0822 & 0.0777 & 0.6346 & 0.9551 \\
		CQFS 30\% & 0.1195 & 0.0782 & 0.0838 & 0.0786 & 0.6388 & 0.9560 \\
		CQFS 40\% & \textbf{0.1228} & \textbf{0.0799} & \textbf{0.0858} & \textbf{0.0817} & 0.6369 & 0.9569 \\
		CQFS 60\% & \textbf{0.1224} & \textbf{0.0797} & \textbf{0.0859} & \textbf{0.0814} & 0.6527 & 0.9567 \\
		CQFS 80\% & \textbf{0.1231} & \textbf{0.0801} & \textbf{0.0860} & \textbf{0.0819} & 0.6585 & 0.9578 \\
		CQFS 95\% & \textbf{0.1234} & \textbf{0.0803} & \textbf{0.0864} & \textbf{0.0824} & 0.6603 & 0.9578 \\
		\bottomrule
	\end{tabular}
\end{specialtable}

Results for CiteULike-a and Xing Challenge 2017 are summarized, respectively, in Tables  \ref{tab:eval:results:cla:hv2} and \ref{tab:eval:results:xing:qpu}. For the sake of clarity, we only report results obtained with CQFS using an ItemKNN CF as the collaborative model, with the rest of the results available in the Supplementary Materials. Both datasets show a different behavior with respect to The Movies Dataset. Feature engineering techniques do not outperform the standard ItemKNN CBF that uses all the features. This suggests that feature engineering may be difficult on these datasets. However, when comparing CQFS with TF-IDF and CFeCBF we notice how, in general, CQFS is more accurate and more robust with different hyperparameters. Because of this, we can say that CQFS is better than the other two methods in giving more importance to a restricted set of useful features.
In particular, CQFS is able to select only 60\% of the features in CiteULike-a keeping a relatively high accuracy and to select 80\% of the features in Xing Challenge 2017 outperforming ItemKNN CBF, CFeCBF and TF-IDF at the same percentage.
In terms of beyond-accuracy metrics we can see how CQFS maintains a consistently high diversity and coverage on CiteULike-a. In Xing Challenge 2017, the coverage obtained by CQFS is higher than in other baselines and close to the one obtained with ItemKNN CBF and all the features. These are important behaviors, since they show that CQFS can select features without compromising on catalog exploration.

\end{paracol}
\nointerlineskip
\begin{specialtable}[H]
    \small
    \widetable
	\caption{Results on CiteULike-a. Presented data can be interpreted as explained in \tabref{tab:eval:results:tmd:hv2}.}
	\label{tab:eval:results:cla:hv2}
	\setlength{\tabcolsep}{3.54mm}
	\begin{tabular}{ccccccccccc}
		\toprule
		\textbf{Models} & \textbf{Precision} & \textbf{Recall} & \textbf{NDCG} & \textbf{MAP} & \textbf{I. Cov.} & \textbf{Gini} & \textbf{MIL} & \boldmath{$\alpha$} & \boldmath{$\beta$} & \boldmath{$s$} \\
		\midrule
		KNN CBF & 0.1638 & 0.2803 & 0.2500 & 0.1735 & 0.9486 & 0.4893 & 0.9937 & - & - & - \\
		\midrule
		TFIDF 40\% & 0.0901 & 0.1443 & 0.1369 & 0.0891 & 0.9210 & 0.4792 & 0.9938 & - & - & - \\
		TFIDF 60\% & 0.1165 & 0.1912 & 0.1742 & 0.1162 & 0.9581 & 0.4654 & 0.9933 & - & - & - \\
		TFIDF 80\% & 0.1348 & 0.2291 & 0.2048 & 0.1375 & 0.9338 & 0.4343 & 0.9927 & - & - & - \\
		TFIDF 95\% & 0.1512 & 0.2545 & 0.2311 & 0.1599 & 0.9668 & 0.5198 & 0.9940 & - & - & - \\
		\midrule
		CFeCBF ItemKNN & 0.1524 & 0.2613 & 0.2319 & 0.1585 & 0.9231 & 0.4485 & 0.9930 & - & - & - \\
		CFeCBF PureSVD & 0.1379 & 0.2323 & 0.2064 & 0.1374 & 0.8976 & 0.3810 & 0.9907 & - & - & - \\
		CFeCBF RP3Beta & 0.1449 & 0.2474 & 0.2205 & 0.1488 & 0.9083 & 0.4186 & 0.9924 & - & - & - \\
		\midrule
		\multicolumn{11}{c}{CQFS trained with ItemKNN} \\
		\midrule
		CQFS 20\% & 0.1504 & 0.2510 & 0.2242 & 0.1529 & 0.9668 & 0.5253 & 0.9943 & 1 & $10^{-4}$ & $10^{2}$ \\
		CQFS 30\% & 0.1522 & 0.2562 & 0.2272 & 0.1549 & 0.9540 & 0.4830 & 0.9935 & 1 & $10^{-4}$ & $10^{2}$ \\
		CQFS 40\% & 0.1535 & 0.2571 & 0.2295 & 0.1571 & 0.9513 & 0.4863 & 0.9936 & 1 & $10^{-4}$ & $10^{2}$ \\
		CQFS 60\% & 0.1606 & 0.2725 & 0.2421 & 0.1665 & 0.9528 & 0.4955 & 0.9939 & 1 & $10^{-4}$ & $10^{2}$ \\
		CQFS 80\% & 0.1602 & 0.2720 & 0.2422 & 0.1662 & 0.9531 & 0.4805 & 0.9936 & 1 & $10^{-2}$ & $10^{2}$ \\
		CQFS 95\% & \textbf{0.1650} & \textbf{0.2816} & 0.2499 & 0.1732 & 0.9567 & 0.4997 & 0.9939 & 1 & $10^{-3}$ & $10^{3}$ \\
		\bottomrule
    \end{tabular}
\end{specialtable}

\vspace{-12pt}

\begin{figure}[H]
\widefigure
    \begin{subfigure}{0.33\linewidth}
        \includegraphics[width=\linewidth]{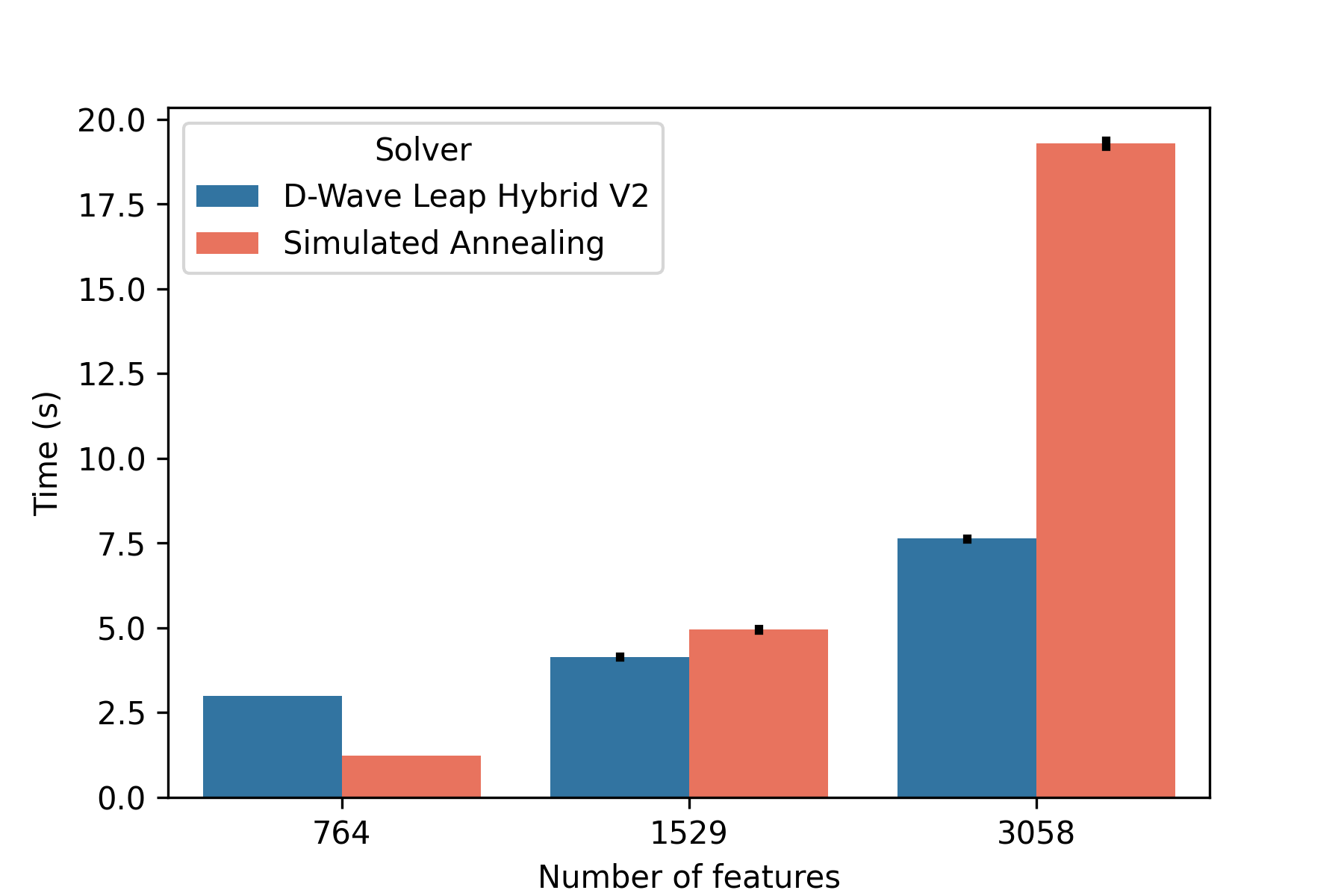}
        \caption{The Movies Dataset}
        \label{fig:scalability:sel:features:tmd}
    \end{subfigure}
    \begin{subfigure}{0.33\linewidth}
        \includegraphics[width=\linewidth]{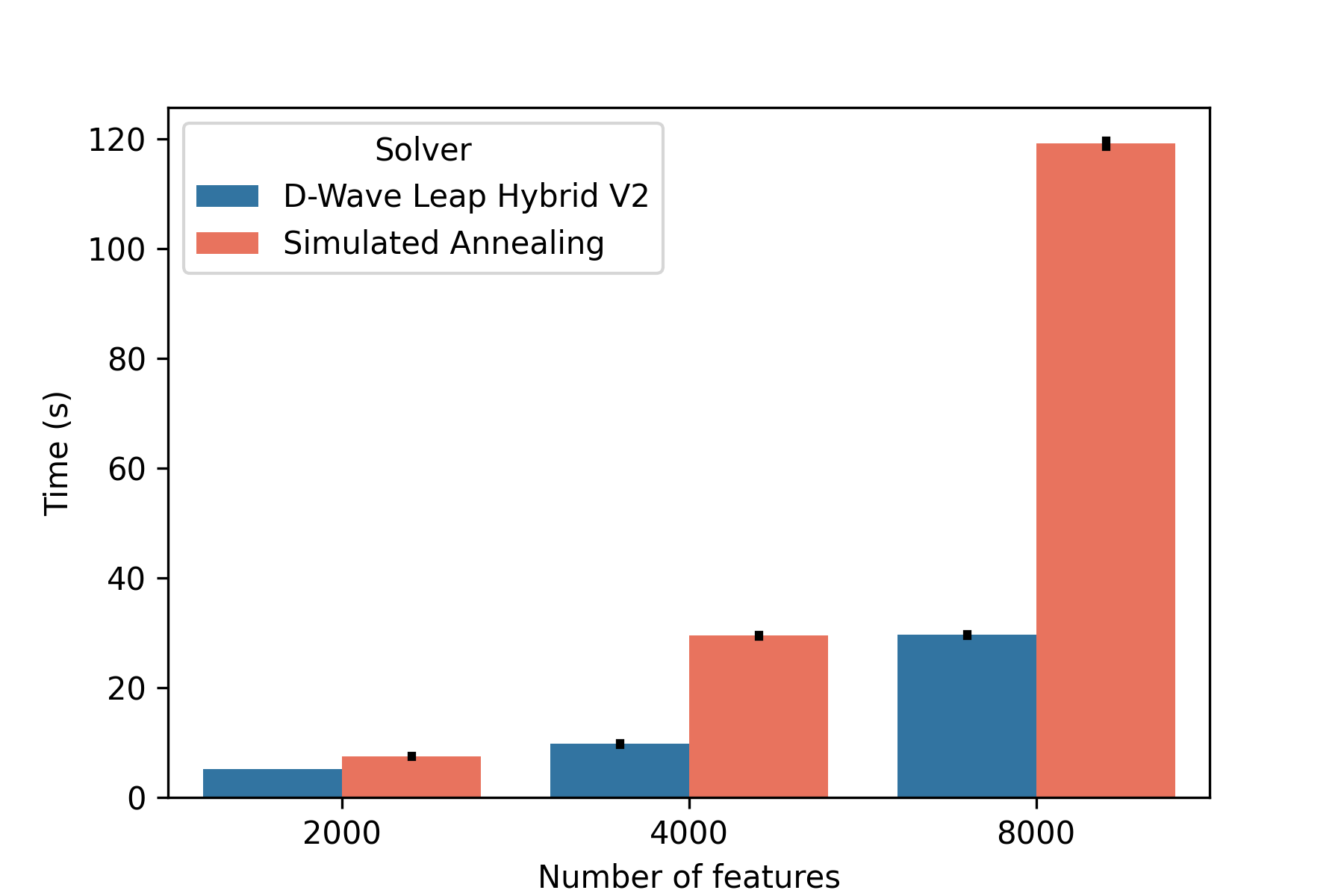}
        \caption{CiteULike-a}
        \label{fig:scalability:sel:features:cla}
    \end{subfigure}
    \begin{subfigure}{0.33\linewidth}
        \includegraphics[width=\linewidth]{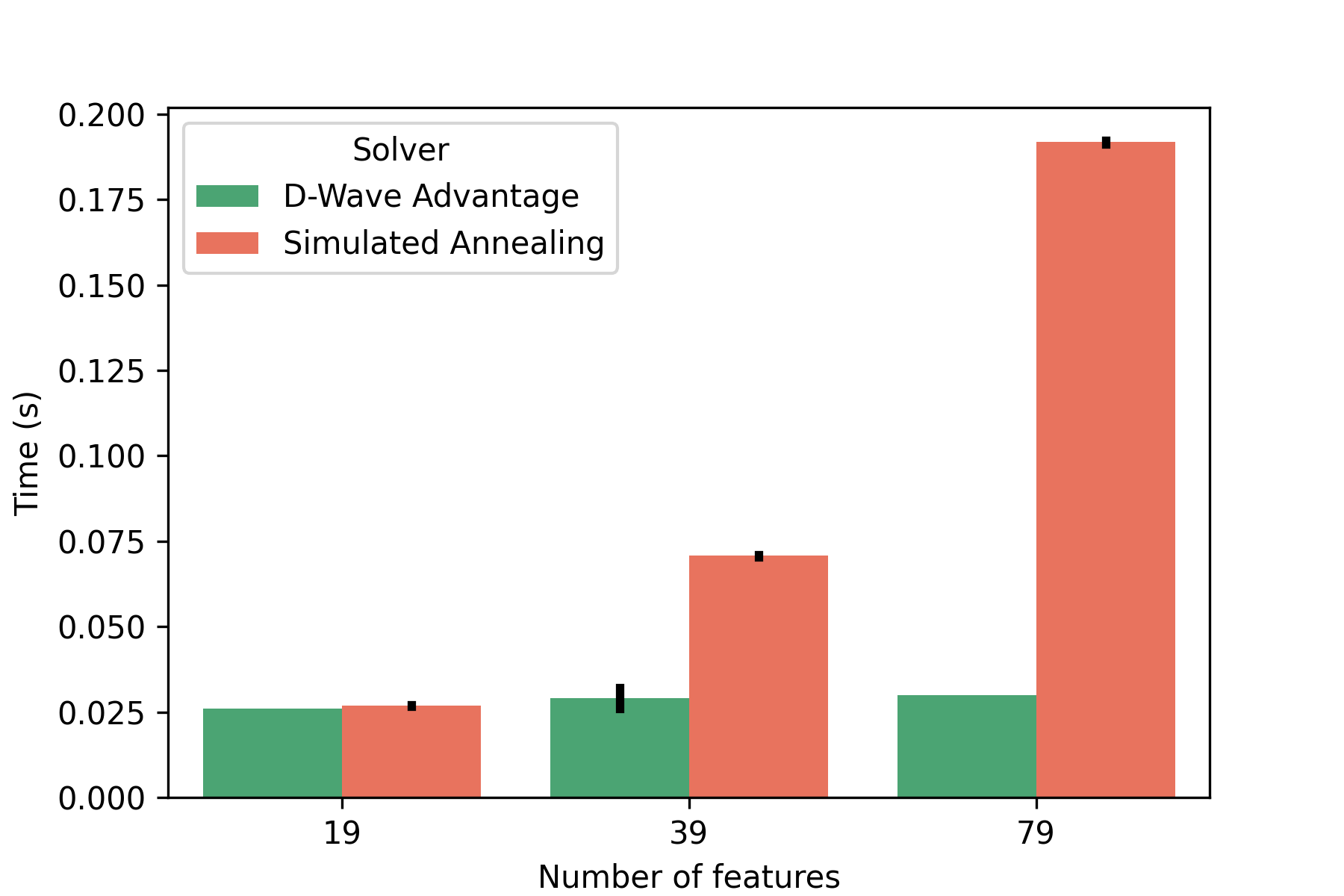}
        \caption{Xing Challenge 2017}
        \label{fig:scalability:sel:features:xing}
    \end{subfigure}
    \vspace{6pt}
    \caption{Scalability analysis on the selection phase, varying feature percentage, comparing {two QUBO solvers, Hybrid QPU and Simulated Annealing for The Movies Dataset (\textbf{a}) 
 and CiteULike-a (\textbf{b}), QPU and Simulated Annealing for Xing Challenge 2017 (\textbf{c}).} While the hybrid quantum solver has a higher or comparable solution time for small instances, it quickly outperforms the SA when the problem size increases. {The QPU, instead, has a constant solution time, independent from the number of features.} The standard deviation is very small.}
    \label{fig:scalability:sel:features}
\end{figure}

\begin{specialtable}[H]
    \small
    \widetable
	\caption{Results on Xing Challenge 2017. Notice that there are no results for CQFS 20\% and 30\% because of the already small number of features in the dataset. Presented data can be interpreted as explained in \tabref{tab:eval:results:tmd:hv2}, apart from the solver used for the CQFS QUBO problem, which in this case is a quantum annealer.}
	\label{tab:eval:results:xing:qpu}
	\setlength{\tabcolsep}{3.54mm}
	\begin{tabular}{ccccccccccc}
		\toprule
		\textbf{Models} & \textbf{Precision} & \textbf{Recall} & \textbf{NDCG} & \textbf{MAP} & \textbf{I. Cov.} & \textbf{Gini} & \textbf{MIL} & \boldmath{$\alpha$} & \boldmath{$\beta$} & \boldmath{$s$} \\
		\midrule
		KNN CBF & 0.0248 & 0.0680 & 0.0525 & 0.0322 & 0.9999 & 0.3572 & 0.9759 & - & - & - \\
		\midrule
		TFIDF 40\% & 0.0013 & 0.0047 & 0.0029 & 0.0016 & 0.2418 & 0.0543 & 0.9597 & - & - & - \\
		TFIDF 60\% & 0.0045 & 0.0147 & 0.0092 & 0.0045 & 0.5253 & 0.1589 & 0.9735 & - & - & - \\
		TFIDF 80\% & 0.0153 & 0.0361 & 0.0250 & 0.0123 & 0.8981 & 0.3004 & 0.9823 & - & - & - \\
		TFIDF 95\% & 0.0297 & 0.0676 & 0.0493 & 0.0287 & 0.9989 & 0.3508 & 0.9874 & - & - & - \\
		\midrule
		CFeCBF ItemKNN & 0.0180 & 0.0489 & 0.0324 & 0.0172 & 0.9865 & 0.3008 & 0.9759 & - & - & - \\
		CFeCBF PureSVD & 0.0009 & 0.0030 & 0.0018 & 0.0009 & 0.7269 & 0.0906 & 0.9735 & - & - & - \\
		CFeCBF RP3Beta & 0.0244 & 0.0596 & 0.0451 & 0.0290 & 0.9985 & 0.3208 & 0.9807 & - & - & - \\
		\midrule
		\multicolumn{11}{c}{CQFS trained with ItemKNN} \\
		\midrule
		CQFS 40\% & 0.0209 & 0.0580 & 0.0454 & 0.0276 & 0.9963 & 0.2851 & 0.9758 & 1 & $10^{-3}$ & $10^{3}$ \\
		CQFS 60\% & 0.0243 & 0.0678 & 0.0513 & 0.0307 & 0.9999 & 0.3424 & 0.9757 & 1 & $10^{-4}$ & $10^{1}$ \\
		CQFS 80\% & 0.0292 & \textbf{0.0744} & \textbf{0.0546} & 0.0314 & 0.9998 & 0.3464 & 0.9765 & 1 & $10^{-3}$ & $10^{2}$ \\
		CQFS 95\% & 0.0241 & 0.0671 & 0.0518 & 0.0314 & 0.9966 & 0.2892 & 0.9822 & 1 & $10^{-4}$ & $10^{3}$ \\
		\bottomrule
    \end{tabular}
\end{specialtable}
\vspace{-4mm}
\begin{paracol}{2}
\switchcolumn


\subsection{Scalability}
\label{sec:scalability}
One of the crucial claims of quantum computing technologies is their ability to provide improved scalability. In order to assess whether the use of a quantum solver proved beneficial in this case we report a scalability analysis.
We carried out these experiments on {all three datasets} and we compared the algorithms in the setting that produced the best results for each. For the baseline algorithms, TF-IDF is reported for $p = 95\%$ and CFeCBF is reported when trained on ItemKNN{ for The Movies Dataset and CiteULike-a and on \rptb for Xing Challenge 2017.} CQFS is reported  when using ItemKNN{,} with $p = 60\%$ for {The Movies Dataset and CiteULike-a and $p = 80\%$ for Xing Challenge 2017.} All the experiments were performed on the same computing resources, apart from Xing Challenge 2017. \added{We used an AWS r5 instance (Intel Xeon Platinum 8000 series with clock up to 3.1 GHz, 16 cores and 128 GB of RAM) and the D-Wave's Leap Hybrid V2 service for The Movies Dataset and CiteULike-a and an AWS r5 instance (48 cores and 384 GB of RAM) and the D-Wave Advantage QPU for Xing Challenge 2017}.
Notice that this is a still developing technology, with early physical limitations. Therefore, we should not be expecting substantial speed-ups against already established techniques on such a generic problem. However, here we show how it is starting to be competitive against classical algorithms.

In \tabref{tab:eval:timecomp} we report the mean and standard deviation of the training time of all the models when trained on the original data with the described setting. \added{For CQFS we report the results of 10 runs, for TF-IDF we report the result of one run and for CFeCBF the time required during the hyperparameter optimization.} 
We can observe that TF-IDF is extremely fast, as expected, as it requires a very simple operation.
CFeCBF, instead, is far more time consuming, requiring in the range of hours for The Movies Dataset{ and Xing Challenge 2017}, with significant variance. Compared to CFeCBF, CQFS is more than \mbox{30 times} faster on The Movies Dataset{,} twice as fast on CiteULike-a{ and 20 times faster on Xing Challenge 2017}, requiring only few minutes.
Notice that CFeCBF has a very high standard deviation, due to its machine learning nature. Indeed, it performs gradient descent with early stopping, which results in CFeCBF having faster runs when it is not learning well.
If we consider the {quantum-based and classical} solvers used for CQFS{,} we can see that the solution time is very close on The Movies Dataset{ and Xing Challenge 2017}, with the SA solver being only marginally slower. On CiteULike-a, instead, there is a bigger difference, given by the higher number of features. It should be noted that the model building phase requires the majority of the time and this limits the gains of the faster solution time allowed by the{ QPU and} Hybrid QPU solver{s}.

In terms of time complexity, the algorithm is dominated by matrix products, with complexity $O(n^3)$.
With respect to the QUBO solving phase, the quantum annealer has a constant solution time of 20 $\upmu$s per sample and the number of samples in our experiments is constant, therefore being $O(1)$. There exist possible strategies to tweak the annealing time as well as considerations on how to select the number of samples depending on the problem size and structure \cite{PhysRevApplied.11.044083}. However, these are complex and mostly open research questions and are therefore out of the scope of this work.
Instead, the hybrid method has a linear complexity by design. The classical SA solver has a higher complexity and in order to provide a better comparison of the {SA solver against the QPU and Hybrid QPU solvers}, we report the solution time of both with various problem sizes: 25\%, 50\% and 100\% of the original features from {the three datasets}.
The construction of the QUBO matrix, \idest model building phase, {and the network time are} not included in this comparison.
\figref{fig:scalability:sel:features} reports the mean and standard deviation of 10 runs of {the classical solver against the quantum-based ones}, comparing only the solution times of an already available QUBO model.
{For The Movies Dataset and CiteULike-a we sampled only one solution for each run, both for the Hybrid QPU and SA solvers, while for Xing Challenge 2017 we sampled 100 solutions for each run, both for the QPU and SA solvers.}
As we can see in \figref{fig:scalability:sel:features}a, on The Movies Dataset the Hybrid QPU solver is slower than SA with a low number of features. However, it rapidly improves the performance, becoming comparable when 50\% of features are used and twice as fast with 100\%.
When compared on CiteULike-a, in \figref{fig:scalability:sel:features}b, we can see how the Hybrid QPU solver gets an even better performance improvement, becoming almost four times as fast when using all the features.
This indicates that the Hybrid QPU solver has a promising scalability.
{In \figref{fig:scalability:sel:features}c we can see that the QPU solver has a constant solution time, independent from the number of features. This is possible because the entire problem fits on the QPU and optimal solutions can be sampled in constant time without hybrid strategies. Due to this, the SA has a solution time an order of magnitude higher than the QPU, when using all the features. Note that the annealing time used by the QPU is a hyperparameter that can be tuned. While it is possible to sample optimal solutions very rapidly on the QPU compared to the classical SA, there is no guarantee on the quality of such solutions. When using the QPU on more complex problems it may become necessary to increase the number of samples, in order to obtain a good solution if those are sampled with low probability. The computational complexity of solving tasks on the QPU will depend on how the number of samples needed to obtain a good solution will grow as the problem becomes bigger and more complex. This is yet another open research question.}
It should be noted that with algorithms having a faster model building phase the advantage of using the{ QPU or the} Hybrid QPU solver{s} will be higher, hence a possible future direction could be reducing the cost of this phase. Furthermore, the scalability of the QPU solver will improve as the hardware becomes able to handle bigger problems, which, if no hybrid algorithm is involved, can be solved by the QPU in constant time, plus the time required for the minor embedding operation.

\end{paracol}
\nointerlineskip

\begin{paracol}{2}
\switchcolumn

\vspace{-12pt}

\begin{specialtable}[H]
	\small
    \caption{Comparison of the training 
 time in seconds for the feature engineering methods on {the three datasets. The SA solver is compared with the Hybrid QPU solver for The Movies Dataset and CiteULike-a, while it's compared with the QPU solver for Xing Challenge 2017.}}
	\label{tab:eval:timecomp}
\setlength{\tabcolsep}{22.2mm}
    	\begin{tabular}{lc}
    		\toprule
    		\multicolumn{2}{c}{(\textbf{a}) The Movies Dataset}\\ \midrule
    		Method & Time (s) \\
    		\midrule
    		TF-IDF 95\% & $ 4.44 \cdot 10^{-3} $ \\
    		CFeCBF & $ \sciexp{7.42}{3} \pm \sciexp{4.09}{3} $ \\
    		CQFS Hybrid & $ \sciexp{2.21}{2} \pm \sciexp{5.95}{-3} $ \\
    		CQFS SA & $ \sciexp{2.33}{2} \pm \sciexp{9.30}{-2} $ \\
    		\midrule
    		\multicolumn{2}{c}{(\textbf{b}) CiteULike-a}\\
\midrule
    		Method & Time (s) \\
    		\midrule
    		TF-IDF 95\% & $ 9.55 \cdot 10^{-3} $ \\
    		CFeCBF & $ \sciexp{8.35}{2} \pm \sciexp{3.41}{2} $ \\
    		CQFS Hybrid & $ \sciexp{4.11}{2} \pm \sciexp{1.15}{-1} $ \\
    		CQFS SA & $ \sciexp{5.00}{2} \pm \sciexp{5.18}{-1} $ \\
    		\midrule
    		\multicolumn{2}{c}{(\textbf{c}) Xing Challenge 2017}\\
    		\midrule
    		Method & Time (s) \\
    		\midrule
    		TF-IDF 95\% & $ 5.77 \cdot 10^{-3} $ \\
    		CFeCBF & $ \sciexp{1.06}{4} \pm \sciexp{3.71}{3} $ \\
    		CQFS QPU & $ \sciexp{5.32}{2} $ \\
    		CQFS SA & $ \sciexp{5.33}{2} \pm \sciexp{5.36}{-4} $ \\
    		\bottomrule
    	\end{tabular}
\end{specialtable}

\vspace{-12pt}

\subsection{Selected Features}
In order to interpret the results for The Movies Dataset, we compare the most weighted features by the baseline algorithms to those selected by CQFS. In \tabref{tab:highest_weighted}a we can see that CQFS does not select any of the 10 most weighted features by TF-IDF, which are mainly production companies. This is due to the fundamentally different strategy the two methods have, with TF-IDF attributing high weight to rarer features while CQFS relying on the user behavior. In \tabref{tab:highest_weighted}b we can see that CQFS instead selects most of the highest weighted features of CFeCBF, mainly movies collections, which is not surprising as both models use collaborative information.

To further interpret the results on The Movies Dataset, we counted how many times each features was selected across all the experiments. In \tabref{tab:eval:sel}a,b we have, respectively, the most and the least selected features by CQFS, with the corresponding selection percentage.
As we can see, CQFS tends to select more informative features, such as movie collections or companies which, upon manual inspection, have shown to be specialized in specific movie genres (such as \emph{Kontsept Film Company}). At the same time, CQFS is able to remove less useful features, such as release years, languages and companies that have been closed for many years (such as \emph{Biograph Company}, closed in 1916).

\begin{specialtable}[H]
	\caption{Features selected by CQFS among the 10 highest weighted features by baseline feature engineering methods.}
	\small
	\label{tab:highest_weighted}
\setlength{\tabcolsep}{16.9mm}
    	\begin{tabular}{lc}
    		\toprule
    		\multicolumn{2}{c}{(\textbf{a}) Comparison with TF-IDF.}\\\midrule
    		Features & Selected \\
    		\midrule
    		Compagnie Cinématographique (CICC) & No \\
    		Clubdeal & No \\
    		Present Pictures & No \\
    		Delphi III Productions & No \\
    		Parallel Films & No \\
    		Carnaby International & No \\
    		Creative England & No \\
    		Coproduction Office & No \\
    		Metro Communications & No \\
    		ARD/Degeto Film GmbH & No \\
    		
    		\bottomrule
    		\multicolumn{2}{c}{(\textbf{b}) Comparison with CFeCBF.}\\\midrule
    		Features & Selected \\
    		\midrule
    		Eon Productions & Yes \\
    		Dragon Ball Z (Movie) Collection & No \\
    		Zatôichi: The Blind Swordsman & Yes \\
    		Friday the 13th Collection & Yes \\
    		James Bond Collection & Yes \\
    		Pokémon Collection & Yes \\
    		The Up Series & No \\
    		Asterix and Obelix (Animation) Collection & No \\
    		Jesse Stone Collection & No \\
    		A Nightmare on Elm Street Collection & Yes \\
    		\bottomrule
    	\end{tabular}
\end{specialtable}

\vspace{-12pt}

\begin{specialtable}[H]
    \small
    \caption{Most and least selected features by CQFS on The Movies Dataset across all the experiments.}
    \label{tab:eval:sel}
\setlength{\tabcolsep}{19.4mm}
        \begin{tabular}{lc}
        \toprule
        \multicolumn{2}{c}{(\textbf{a}) Most selected features.}\\\midrule
                                                     Feature &   Percentage \\
        \midrule
                                   Transformers Collection &  97.72\% \\
                                        K/O Paper Products &  97.70\% \\
                                     Kontsept Film Company &  97.62\% \\
                            Mission: Impossible Collection &  95.62\% \\
                                                  Chaocorp &  95.20\% \\
                                  Laura Ziskin Productions &  94.97\% \\
                                 The Terminator Collection &  94.88\% \\
                                          X-Men Collection &  94.88\% \\
                              Parkes/MacDonald Productions &  94.80\% \\
                                     The Bourne Collection &  94.80\% \\
                                     \midrule
                                     \multicolumn{2}{c}{(\textbf{b}) Least selected features.}\\\midrule
                                                     feature &   Percentage \\
        \midrule
                                                      1898 &  37.53\% \\
                                          Biograph Company &  37.94\% \\
                                    PM Entertainment Group &  38.25\% \\
                                             Lippert Films &  38.70\% \\
                                                Flora Film &  39.19\% \\
                                                      2001 &  40.00\% \\
                                                      1964 &  40.00\% \\
                                                   Western &  40.00\% \\
                                   Spoken language svenska &  40.00\% \\
                                                      1977 &  40.00\% \\
        \bottomrule
        \end{tabular}
\end{specialtable}

\section{Conclusions}
\label{sec:conclusions}
In this paper, we have shown an application of quantum annealing to recommender systems by proposing a hybrid feature selection approach for the item cold start problem.
Although quantum annealing is still an emerging technology, with some technological limitations, we are already starting to see advantages in some cases.
Indeed, depending on the dataset and the collaborative model used in the training phase, the proposed algorithm is able to provide competitive recommendation quality even when selecting as little as 30\% of the original features.
Moreover, across all the experiments we can see how selections made with CQFS result in recommendations with better diversity and accuracy with respect to other feature engineering methods.
The proposed algorithm can be solved with both quantum and classical solvers in comparable time.
However, the Hybrid QPU solver shows promising scalability for bigger instances.
Overall, our results indicate that quantum annealers are becoming viable solvers for applied research.

\vspace{6pt}

\supplementary{The following are available online 
 at \url{www.mdpi.com/xxx/s1}, supplementary materials provided with this work comprise: Table S1: Complete CQFS results on CiteULike-a. Table S2: Complete CQFS results on Xing Challenge 2017. \deleted{A version of the CiteULike-a dataset already compatible with our software. }}

\authorcontributions{Conceptualization, \imf and \ipc; methodology, \irn and \imf; software, \irn; investigation, \irn; resources, \imf and \ipc; validation, \irn and \imf; writing---original draft preparation, \irn; writing---review and editing, \imf and \ipc; visualization, \irn; supervision, \imf and \ipc All authors have read and agreed to the published version of the manuscript.}

\funding{\added{We acknowledge the CINECA award under the ISCRA initiative, for the availability of high performance computing resources and support.
}}

\dataavailability{Three dataset were used for this work.
The Movies Dataset is a dataset publicly available on Kaggle at 
 \url{https://www.kaggle.com/rounakbanik/the-movies-dataset} (accessed on 27 July 2021).
CiteULike-a is a dataset from the site CiteULike, available at \url{https://github.com/js05212/citeulike-a} (accessed on 27 July 2021).
We also provide a version of this dataset already compatible with our software in the supplementary materials.
Xing Challenge 2017 is a private dataset from a previous ACM RecSys Challenge. It is no longer available, as of now.}

\conflictsofinterest{The authors declare no conflict of interest.}

\end{paracol}

\reftitle{References}













\end{document}